\documentclass{article}

\usepackage{arxiv}

\usepackage[utf8]{inputenc} 
\usepackage[T1]{fontenc}    
\usepackage{setspace}
\usepackage{hyperref}       
\usepackage{url}            
\usepackage{booktabs}       
\usepackage{amsfonts}       
\usepackage{nicefrac}       
\usepackage{microtype}      
\usepackage{lipsum}		
\usepackage{graphicx}
\usepackage{natbib}
\usepackage{doi}

\usepackage{acro}

\usepackage{subcaption}

\usepackage{algorithm} 
\usepackage{algorithmic}  
\usepackage[algo2e]{algorithm2e}

\usepackage[acronym]{glossaries}
\usepackage{booktabs}

\usepackage{wrapfig}
\usepackage{textcomp}

\usepackage{amsmath} 
\usepackage{setspace}

\usepackage{placeins} 

\usepackage{enumitem}

\newcommand{\eg}{e.g., }
\newcommand{\ie}{i.e., }

\newcommand{\etc}{etc. }

\newcommand{\kmeans}{\textit{K-means }}
\newcommand{\drkid}{Dr.\ KID }

\DeclareAcronym{cvt}{
  short=CVT,
  long=Centroidal {V}oronoi tessellation,
}

\title{Dr.\ KID: Direct Remeshing and K-set Isometric Decomposition for Scalable  Physicalization of Organic Shapes}
 
\author{%
    \href{https://orcid.org/0000-0001-5864-1888}{\includegraphics[scale=0.06]{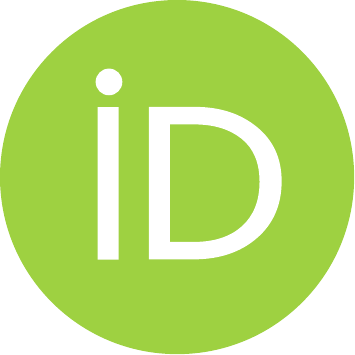}\hspace{1mm}Dawar Khan},
    \href{https://orcid.org/0000-0002-9015-2897}{\includegraphics[scale=0.06]{figures/orcid.pdf}\hspace{1mm}Ciril Bohak}, and
    \href{https://orcid.org/0000-0003-4248-6574}{\includegraphics[scale=0.06]{figures/orcid.pdf}\hspace{1mm}Ivan Viola}
    \\
	Visual Computing Center,
	King Abdullah University of Science and Technology,\\
	Thuwal, Kingdom of Saudi Arabia \\
	\texttt{\{dawar.khan, ciril.bohak, ivan.viola\}@kaust.edu.sa} \\
}

\date{}

\renewcommand{\shorttitle}{Dr.\ KID for Scalable Physicalization of Organic Shapes}

\hypersetup{
pdftitle={Dr.\ KID: Direct Remeshing and K-set Isometric Decomposition for Scalable  Physicalization of Organic Shapes},
pdfsubject={q-bio.NC, q-bio.QM},
pdfauthor={Dawar Khan, Ciril Bohak, Ivan Viola},
pdfkeywords={Physicalization, Physical visualization, 3D printing, Isometric decomposition, Direct remeshing, Biological structures, Intracellular compartments.},
}

\begin{document}
\maketitle

\begin{figure}[!h]
    \centering
\vskip -2.0cm
  \includegraphics[width=\linewidth]{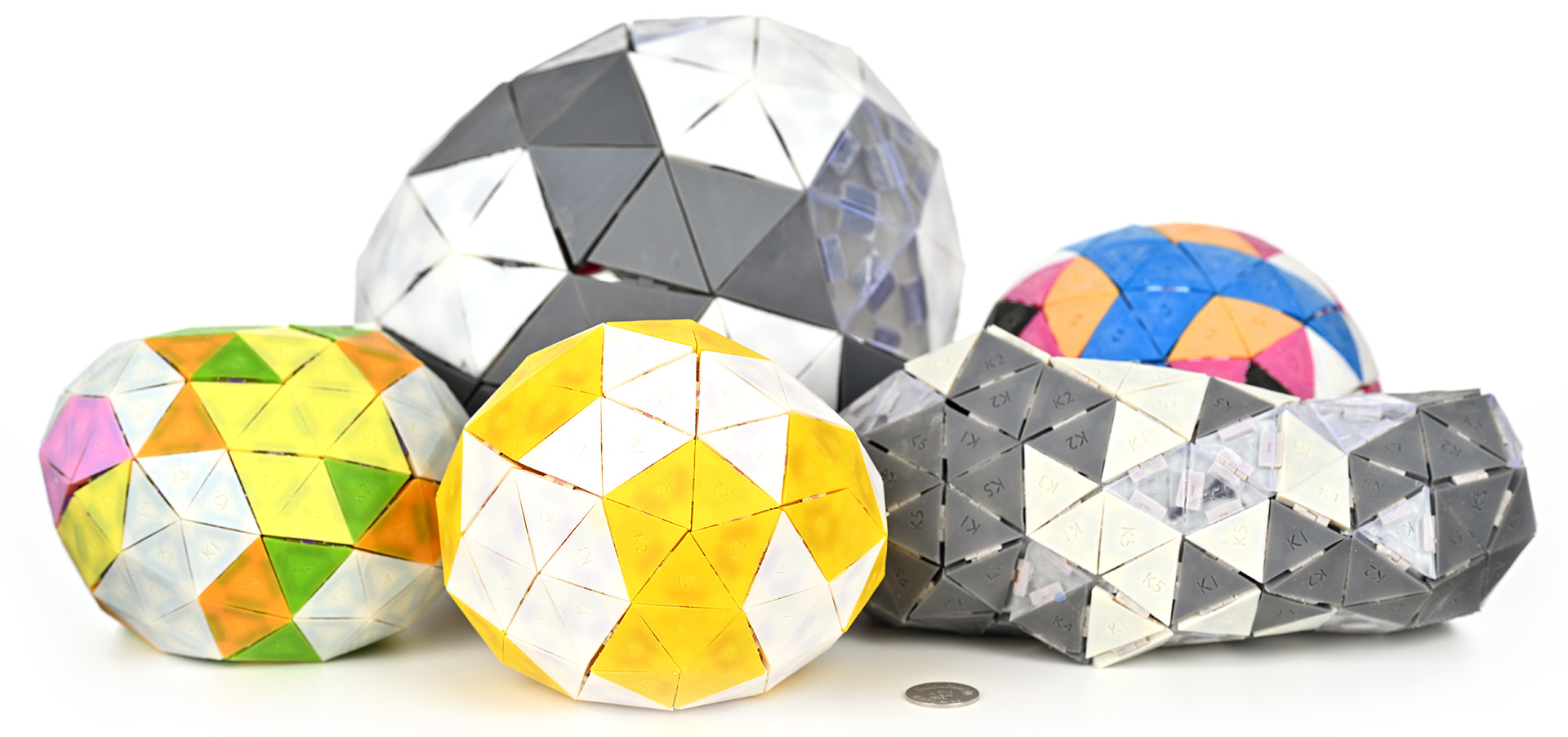}
\vskip -0.35cm
  \caption{%
    The physicalization of potato-shaped biological structures with $k$ types of triangles. Back row: SARS-CoV-2 virion membrane (left)  with $k=2$, SARS-CoV-2 virion membrane with smooth triangle patches (right),  using $k=6$, and front row: cell nuclei membrane (left),  using $k=5$, SARS-CoV-2 virion membrane (center),  using $k=2$, mitochondria outer membrane (right),  using $k=6$.
  }
  \label{fig:teaser}
\end{figure}

\begin{abstract}
\drkid is an algorithm that uses isometric decomposition for the physicalization of potato-shaped organic models in a puzzle fashion. The algorithm begins with creating a simple, regular triangular surface mesh of organic shapes, followed by iterative \kmeans clustering and remeshing. For clustering, we need similarity between triangles (segments) which is defined as a distance function. The distance function maps each triangle's shape to a single point in the virtual 3D space. Thus, the distance between the triangles indicates their degree of dissimilarity. \kmeans clustering uses this distance and sorts segments into \textit{k} classes. After this, remeshing is applied to minimize the distance between triangles within the same cluster by making their shapes identical. Clustering and remeshing are repeated until the distance between triangles in the same cluster reaches an acceptable threshold. We adopt a curvature-aware strategy to determine the surface thickness and finalize puzzle pieces for 3D printing. Identical hinges and holes are created for assembling the puzzle components. For smoother outcomes, we use triangle subdivision along with curvature-aware clustering, generating curved triangular patches for 3D printing. Our algorithm was evaluated using various models, and the 3D-printed results were analyzed. Findings indicate that our algorithm performs reliably on target organic shapes with minimal loss of input geometry.


    
\end{abstract}


\section{Introduction}
\maketitle
\label{sec:intro}
Physical models are powerful tools for understanding and comprehending complex objects and systems. They provide a simplified representation of an object's shape, structure, function, and inner composition, making it easier for individuals to grasp the intricacies of the object~\citep{Bailey1998,Casas2015}. Physical visualization bridges the gap between digital data and its physical composition~\citep{Bader2018}. This is especially true for objects that are too large to take in all details at once, too small to depict their inner structure, or too complex to fully understand the interconnectedness of their parts and how they work together. Physical visualizations are often implemented as 3D puzzles.

3D puzzles, in particular, offer an engaging and challenging form of entertainment that can help improve spatial reasoning and problem-solving skills~\citep{Verdine2014,Nicolaidou2021,Mei-Ling2022}. They illustrate the real-world object's inner structure in a way that can be easily understood, allowing individuals to gain a deeper understanding of the object and its workings. In addition to being a source of entertainment, 3D puzzles can also be used in education as a tool to teach how things are built and how they function~\citep{Casas2015}. They can also be used in outreach to attract people's attention and ignite an interest in a specific topic. For such use cases, puzzles are actually data physicalization~\citep{Moere2008,Zhao2008,Jansen2013,Jansen2015,Dragicevic2020,Djavaherpour2021} intended to familiarize the users with the data they are based on.

A world of mesoscale biological models is an appropriate domain for such puzzles representing the intricate and detailed structures of living organisms at the cellular and subcellular levels. Mesoscale biological structures are typically very complex and are built of many small building blocks. Assembling such puzzles allows individuals to gain a deeper understanding and appreciation of biological systems' complex and dynamic nature. Furthermore, the physical nature of these puzzles provides for a hands-on learning experience, helping to make the information more memorable and engaging~\citep{Echavarria2020}, which is beneficial in educational settings as a tool for teaching anatomy, physiology, and biology~\citep{Hidayat2019}. They can also be beneficial for research purposes, for example, to help scientists study the structure of proteins or viruses~\citep{Nguyen2021}.



For a 3D puzzle to be suitable for injection molding production, it is important to minimize the number of different parts in the design because the cost of producing the molds for injection molding is directly related to the number of elements in the puzzle and is the costliest part in the pipeline. It is also important to keep in mind that the puzzle's design should be optimized for the injection molding process to ensure that it can be produced efficiently and with minimal defects. This could include simplifying the puzzle's geometry, avoiding deep undercuts and sharp corners, and ensuring that the puzzle's parts can be easily removed from the mold.

To enable scalable generation of 3D puzzles representing biological systems such as viruses, organelles, or cells, it is meaningful to use the molding process for generating all the building blocks. Often, these \emph{potato-shaped} structures are surrounded by a membrane formed by a lipid bilayer. The membrane is richly decorated with macromolecular protein complexes, which are also forming the lumen of the biological system. Additionally, there are numerous fibrous structures, such as microtubules, actin fibers, and genetic macromolecules, that contribute to the internal ultrastructure of the biological system. In a viral particle, for example, a few unique protein structures form the mature virus, which are instantiated in the virus numerous times. This is ideal for the molding process for the purposes of data physicalization. One of the main problems is how to represent the lipid bilayer, which is the basis of the shape of the biological system. In the case of an ideal sphere, the surface can be assembled from identical spherical triangular patches. However, real biological structures are never perfect spheres. Their shapes are more similar to a potato, cucumber, or bean. Tesselating the surface of such 3D shapes results in a set of unique triangles, \ie every triangle is different. Such tessellation is prohibitive in the context of scalable 3D puzzle generation, as the number of unique  pieces is too high for viable puzzle production. Therefore, we need a tessellation that approximates the shape well and, at the same time, is made out of very few distinct classes of triangle patches. This problem constitutes the intellectual challenge of this paper. 

We present a puzzle-generation system called \drkid, which generates a surface mesh of biological structure and decomposes it into reconfigurable puzzle components (segments). The segments can be either planar triangles (planar patches) or curved triangular patches (curved patches). We focus on the isometric decomposition of the surface, where we take the surface mesh of the model and decompose it into a set of $k$ isometric segments (identical components). Following the surface thickening stage, we create connector structures for assembling and disassembling the puzzle segments. We use unique hinge joint connector parts for connecting the segments. For these connector hinge joints, we created identical holes on each side of all the triangle segments. \drkid solves a geometric problem and provides a real-world prototype for scientific outreach using data physicalization. It is scoped for depicting a micron-sized biological system but can be used for creating 3D puzzles of any potato-like shapes.

The technical contributions of the proposed work are summarized below: 
\begin{itemize}[nolistsep]
    \item A new distance measure for \kmeans isometric decomposition of the surface into triangular patches.
    
    \item Surface remeshing with novel cluster-aware local operators for within-cluster distances minimization and thus improving isometries of within-cluster patches.

    \item A novel automatic process of curvature-aware surface thickening of triangular patches and connector-placement design.

    \item Application of isometric decomposition for data physicalization of potato-shaped objects showcased on biological structures.
    
    
\end{itemize}

\section{Related Work}
\label{sec:related-work}
This section presents related work from various domains, including scalable physicalization, Isometric decomposition, 3D puzzles, and surface remeshing.

Isometric decomposition of a surface is a challenging task that can generate identical puzzle parts for reconfigurable 3D models. In addition to 3D puzzles, isometric decomposition plays an important role in various fields of computational design, including architectural geometry, fabrication, modeling, surface reconstruction, and more. One of its main uses is generating isometric segments of the input surface. This complex geometric problem enhances reusability, reducing both complexity and costs~\citep{Jiang2021,Zimmer2014a,Huard2015,Liu2021}. Isometric decomposition has applications in tiling~\citep{Fu2010}, modeling, and fabrication~\citep{Liu2021}. Depending on the surface mesh and application, surfaces can be decomposed into triangles~\citep{Liu2021,Singh2010}, quads~\citep{Fu2010}, or other polygonal segments. 

To the best of our knowledge, the first attempt at the isometric decomposition of curved surfaces was conducted by~\cite{Singh2010}. They employed a set of template triangles called canonical triangles and remeshed the model to make each triangle in the mesh identical to one of those. Their method~\citep{Singh2010} does not allow topology change and needs higher numbers of triangles to preserve the desired shape. Furthermore, they start with a single cluster and add more clusters later. They used global optimization instead of direct remeshing. Although their method accommodates curved surfaces, the canonical triangles remain planar, and there is no mechanism to address surface thickness. Planar penalization~\citep{Huard2015} is another attempt to reduce complexity by creating repetitive patterns. The work by~\cite{Fu2010} on K-set tileable surfaces presents a similar approach for quad meshes, generating similar quads. This approach offers a fascinating solution for minimizing the number of fabricated components into a given number of $k$ quads. However, using quad meshes can drastically alter the original shape when minimizing the value of $k$. While the authors mention that this idea can be extended for puzzle-like reassembleable applications, no such extension currently exists.

\cite{Liu2021} proposed a method for modeling and fabrication with a minimal number of classes of equivalent triangles. This method is the most relevant for our current work. However, this method uses existing template triangles, which are meant for large fabrication models and therefore do not require addressing their thickness. The connection among the triangles is established via holes and nylon cable. Moreover, the triangles are planar, and the curvature is created at connection points among the triangles. Therefore, the smoothness of the models is not realistic.

In summary, the literature review yields several methods for isometric decomposition. However, due to their limitations, these methods are not practical for puzzles and physicalization. For example, all these methods anticipate having existing templates rather than supporting a dynamic number of pieces. The structures generated with them also rely on external support to keep the individual surfaces in place. Furthermore, the existing methods have no mechanisms for surface thickness, which is essential for the independent assembly of the models or surface smoothness. The k-set tileable quads method by~\cite{Fu2010} drastically changes the input model if decreasing the number of segments. Moreover, there is less attenuation toward biological models. Unlike CAD models or general architectural geometry, biological models are more challenging.

Isometric patterns are highly encouraged in architectural geometry and civil engineering. In this regard,~\cite{Jiang2021} used isometric bending of surfaces via a small set of molds to create  manufacturable tiles, which addresses the problem of representing free forms. They only used constant Gaussian curvature for this paneling. Developability of a B-spline surface~\citep{Gavriil2019} improves the paneling task and can reduce manufacturing costs significantly. This paneling method~\citep{Gavriil2019} locally approximates the 1-dimensional Gauss image with a circle. The method gives smoother results with higher efficiency. However, it is limited to the grid-like panel arrangement.

\textit{Zometool}\footnote{\url{https://zometool.com/}} is a mathematical and molecular modeling kit, which is a popular visualization and physicalization tool with a specific focus on repetitive patterns such as molecular structures, crystal lattices, and mathematical constructs. Further research~\citep{Zimmer2014a,Zimmer2014b}  extended the scope of \textit{Zometool} toward free-form, disk-topology surfaces. The studies~\citep{Zimmer2014a,Zimmer2014b} provide an insight into utilizing \textit{Zometool} for architectural applications. Adopting the advancing front strategy, they start from a single vertex and grow forward for the physicalization of the model. They use hybrid meshes (mixed with quads and triangles) for surface representation. They use nine different types of edges. Users, however, cannot select the number of classes. Therefore, there is no option to minimize the number of used polygon classes.

3D puzzles captivate the imagination with their wide-ranging and intricate designs. They encompass an array of types, such as the essential 2-manifold jigsaw puzzles that form the surfaces of 3D objects, as highlighted by~\cite{Coffin2006}. Polyomino puzzles, presented by~\cite{Lo2009}, reminiscent of intricate Tetris-like shapes, come together to create elaborate 3D objects. Burr puzzles delight with their interlocking pieces that, when assembled, reveal complete 3D models, as presented by~\cite{Xin2011}. The enchanting world of recursively interlocking puzzles, presented by~\cite{Song2012}, takes the complexity of burr puzzles to a higher level. Dissection puzzles offer a transformative experience, as they can be reassembled into various forms as presented by~\cite{Sequin2012}. Twisty puzzles invite engagement through their assembly or disassembly via twisting motions, as shown by~\cite{Sun2015}. \cite{Elber2022} present some recent improvements, which address the 3D jigsaw puzzles over 2-manifolds, while~\cite{Tang2019} present a novel approach to the computational design of 3D dissection puzzles.

Fabricated 3D puzzles created based on scientific data exemplify data physicalization, transforming abstract information into tangible objects~\citep{deFreitas2022}. Tangible data visualization facilitates a more intuitive, engaging, and immersive exploration of complex data structures~\citep{Eslambolchilar2023,Djavaherpour2021}. As users interact with the printed pieces and strive to solve the puzzles, they actively decipher the encoded data, enhancing their cognitive understanding and promoting deeper insights. Consequently, 3D printed puzzles, as a manifestation of data physicalization, offer an innovative and accessible means of visualizing, analyzing, and comprehending information, allowing users to connect with the data on a visceral level, transcending the limitations of traditional, screen-based representations.

The fabrication of 3D puzzles varies greatly depending on the puzzle type. For different puzzle types, multiple factors must be considered, such as printing 3D objects that exceed the printer's working volume and can later be assembled using custom connectors~\citep{Luo2012, Yao2015}, need to be optimally packed~\citep{Chen2015}, or designed as interlocking parts~\citep{Song2015}. Some approaches also approximate 3D models using multiple planes, simplifying the fabrication process to two dimensions~\citep{Chen2013}, and utilizing unique fabrication methods for the base structure and detailed sections, which can then be merged into one piece~\citep{Song2016}. Although cost optimization is most significant in large-scale applications like architectural design~\citep{Eigensatz2010}, it offers considerable savings by incorporating 3D printing into the injection molding process chain~\citep{Tosello2019}.

Mesh processing has always been an effective way to represent surfaces~\citep{Lorensen1987}, allowing a variety of surface analyses. They can provide a base for accurate surface decomposition. Surface remeshing~\citep{Khan2022,Alliez2008} alters the meshes for their quality improvement and/or some other remeshing objectives. Remeshing algorithms are either based on global optimization or use selected local operators. \ac{cvt}~\citep{Lloyd1982,Liu2009} has been widely used for surface remeshing~\citep{Khan2022}. It is an improved type of Voronoi diagram, which relocates each seed to  the mass center of its Voronoi cell. Typically, this relocation is achieved by minimizing the specific energy function.
 
Typically, isometric decomposition algorithms have two main tasks, pattern matching to find the similarity among the patches and remeshing to make the patches similar. Pattern matching quantifies the degree of similarity among the patches, whereas remeshing alters their shapes to make them identical to other shapes in the class. Mesh representation also provides opportunities for surface smoothing by triangle subdivision~\citep{Loop1987}. We refer the readers to two comprehensive survey articles on surface remeshing~\citep{Khan2022,Alliez2008}.
\section{K-set Isometric Decomposition}
\label{sec:method}
 The domain of our algorithm is a family of biological structures having a potato-like shape (\eg mitochondria, lysosome, endosomes, cellular nuclei, virus particles etc.). The goal is to create a cost-effective and appealing physicalization of these shapes. To achieve this goal, we address two challenges: (1) including isometric decomposition with geometric fidelity and (2) curvature-aware modeling of decomposed parts as assemblable parts. The problem is formally stated as follows:
 
 \paragraph{Problem statement:} Our goal is to decompose the surface mesh $M_i$ of these structures into a given number of $k$ classes of mesh segments $C=\{c_1, c_2, c_3, \ldots, c_k\}$. Each class $c_i$ contains a finite number of mesh segments $P=\{p_i\}_{i=1}^{|c_i|}$ satisfying the following requirement conditions: 
\begin{itemize}[nolistsep]
    \item \textit{Similarity and distinctions}: Any two patches belonging to the same class should be identical to each other and different from patches from other classes. Mathematically, if $p_i \in c_i$ and $p_j \in c_i$, then both $p_i$ and $p_j$ must be identical and distinct otherwise.
    
    \item \textit{Planar vs.\ curved segments}: Depending on the user's choice, the patches can be planar (triangular patches) or curved mesh segments (curved patches).
    
    \item \textit{Geometric fidelity and approximation}: The segmented mesh (collection of all patches) should preserve the input shape with an acceptable approximation error.
    
    \item \textit{Reconfigurable objects}: The patches are linked with identical connectors and holes, yielding a set of reconfigurable 3D objects (to be manufactured for the 3D puzzle).
    
    \item \textit{Patch thickness}: Such reconfigurable patches require thickness which is, in our case, provided by the user.
\end{itemize}   

\begin{figure}[tb]
    \centering
    \includegraphics[width=0.8\linewidth]{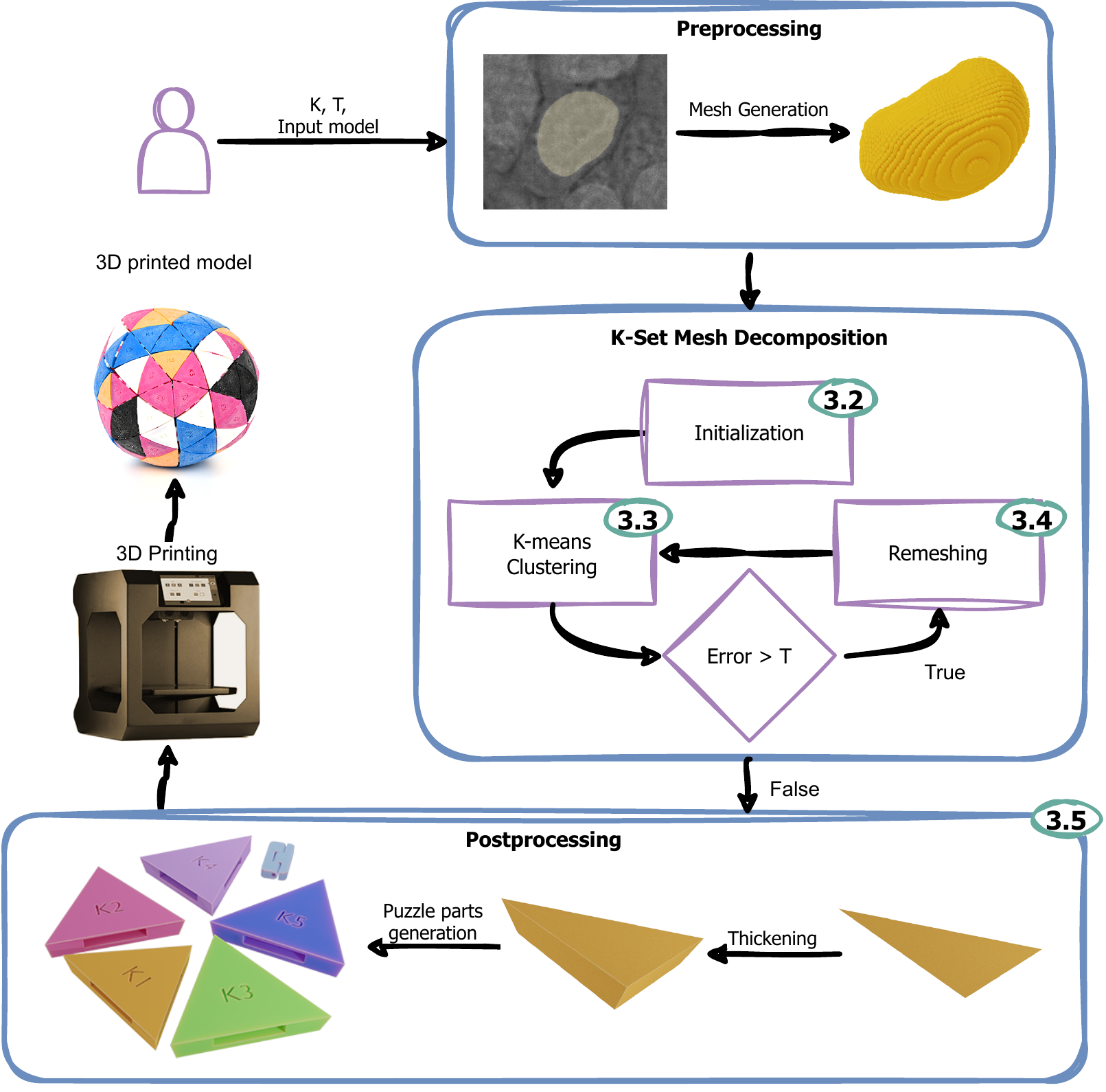}
    \caption{The overview of the presented method illustrating inputs and outputs of the individual steps.}
    \label{fig:overview}
\end{figure}
  
\subsection{Method Overview}
\label{sec:algoOverview}
The method overview is presented in \autoref{fig:overview} and in more detail by \autoref{alg:main}, which presents a pseudocode for the overall physicalization, starting with the 3D shape and ending with the puzzle pieces. The algorithm takes input mesh ($M_i$) generated from the input 3D structure, error threshold (T), which represents the acceptable tolerance during remeshing, and the value of $k$ representing the number of classes. The threshold (T) is the maximum permissible error for k-set isometric decomposition. It provides a convergence point (stopping point) to our iterative algorithm. The algorithm starts with mesh generation and initialization (\autoref{sec:init}). The overall idea is to remesh $M_i$ and get a final mesh $M_f$ satisfying the aforementioned conditions. For clustering, we use a distance measure that maps the similarity among triangles into square Euclidean distances in a 3D space. For K-set isometric triangulation, we minimize the distance defined by \autoref{main_eq} by iteratively executing the two consecutive steps: (1) K-means clustering (\autoref{sec:K-means}) and (2) remeshing (\autoref{sec:remesh}). Clustering, remeshing, and updating each triangle's position in our 3D space are repeated until the algorithm converges. After this, the triangles are processed for connecting structures consisting of a hinge joint (connector) and corresponding holes in individual parts (holes). We present an automatic module for making curvature-aware thickness and connectors and holes.

The default classification is for planar triangular patches. However, we also support smoother curved triangular patches by applying the Loop subdivision~\citep{Loop1987}, which gives us a smoother surface (see \autoref{sec:subdivision}). Again, we apply patch-wise classification with a higher value of $k$ using \kmeans with curvature information to get a final classification of the curved patches.

\begin{figure}[tb]
    \centering
    \includegraphics[width=\linewidth]{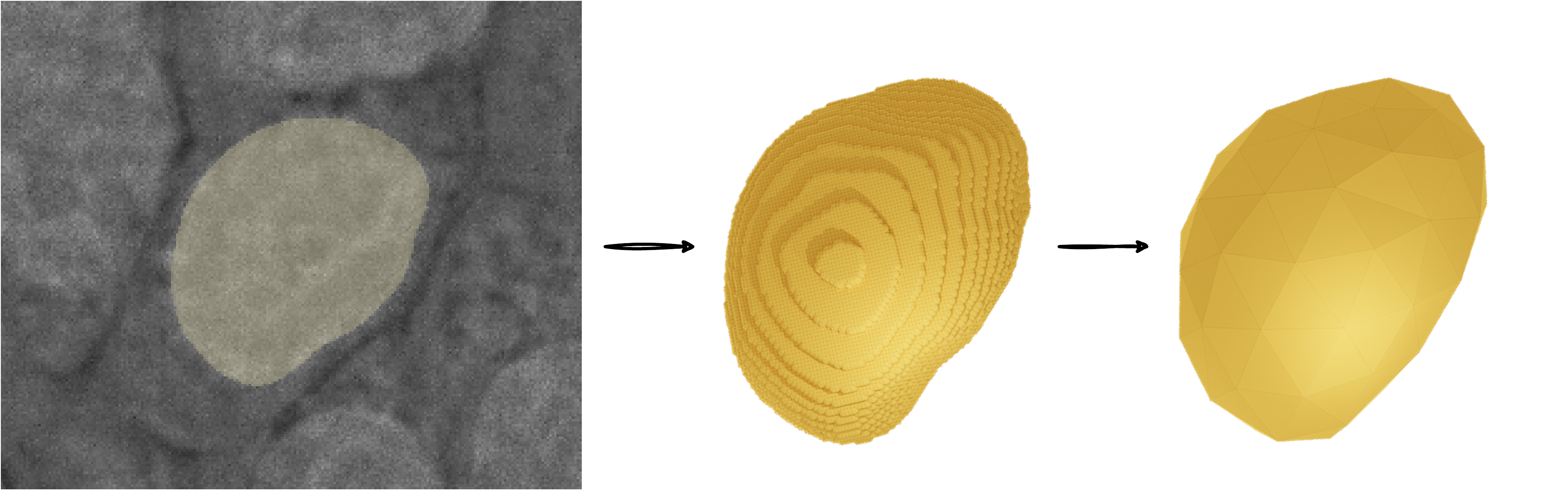}
    \caption{Converting segmented volumetric data (left) to mesh representation using Marching Cubes (middle) and mesh refinement using \ac{cvt} (right).}
    \label{fig:data-preparation}
\end{figure}
\begin{algorithm}
\caption{K-Set Isometric Decomposition}
\label{alg:main}
\KwData{$M_i, T, k, 3DShape$}
\KwResult{$M_f$, k types identical thickened triangles and connectors}
  \begin{algorithmic}[1]
    \STATE $M_i \gets \mathrm{SurfaceMesh}(3DShape)$
    \STATE $M_s \gets \mathrm{Init}(M_i)$
    \STATE $M_f \gets \mathrm{Init}(M_s)$
    \STATE $Virtual3Dspcae \gets \mathrm{CalculatePositionin3Dspace}(M_f)$
    \STATE $Convergence \gets False$, $Itr \gets1$
    \WHILE{$(!Convergence)$}
      \STATE $\left[Labels, WithinClusterDistances\right] \gets \mathrm{Clustering}(M_f, k, T)$
      // Calling~\autoref{alg:clustering}
      \STATE $M_f \gets \mathrm{UpdateClusteringLabels}(Labels,M_f)$
      \IF{$(T\ge \max(WithinClusterDistances))$}
        \STATE $Convergence \gets True$ // (see~\autoref{maxCError_eq} for convergence point)
      \ELSE
        \STATE $M_f \gets Remeshing(M_s, k, T, Itr++)$ // Calling~\autoref{alg:remesh}
        \STATE $M_f \gets \mathrm{UpdatePositionin3Dspace}(M_f)$
      \ENDIF
    \ENDWHILE
    \STATE $VNormals \gets \mathrm{VerticesNormals}(M_f)$ // for surface thickness.
    \STATE $\mathrm{GenerateThickenedTriangles}(M_f)$
    \STATE \textit{Make connector-placement and Hinges.}
    \STATE $END$
  \end{algorithmic}
\end{algorithm}

\subsection{Mesh Generation and Initialization }
\label{sec:init}
For input, we take segmented volumetric data of biological structures such as mitochondria, viral virions, intracellular compartments \etc The segmented structures are converted from voxelized to mesh representation (see \autoref{fig:data-preparation}). We use the Marching cubes algorithm~\citep{Lorensen1987} for generating our first raw mesh $(M_i)$ from the 3D structure. However, this raw mesh has several defects, including self-intersections, higher complexity, and redundant elements. Therefore, prior to actual classification and remeshing, careful refinement is required. We refine $M_i$ in initialization step using \ac{cvt}~\citep{Lloyd1982,Liu2009}. The initialization step not only improves mesh quality but also simplifies the mesh by setting the number of seeds in \ac{cvt}. This refined and simplified mesh is denoted by $M_s$ throughout this paper. More specifically, the raw mesh is remeshed with two different implementations of the \ac{cvt}, including 5 iterations of the Lloyd algorithm~\citep{Lloyd1982} followed by 30 iterations of the quasi-Newton optimization~\citep{Liu2009}. With \ac{cvt} initialization, we also specify the number of vertices for the surface mesh, through which we can simplify mesh (reduce the total number of vertices). \ac{cvt}~\citep{Lloyd1982,Liu2009}, being efficient and easy to implement, has been widely used in surface remeshing and related research~\citep{Khan2022}. It simply relocates each seed toward the center of mass of the Voronoi cell. 

\begin{figure}[tb]
    \centering
    \includegraphics[width=\linewidth]{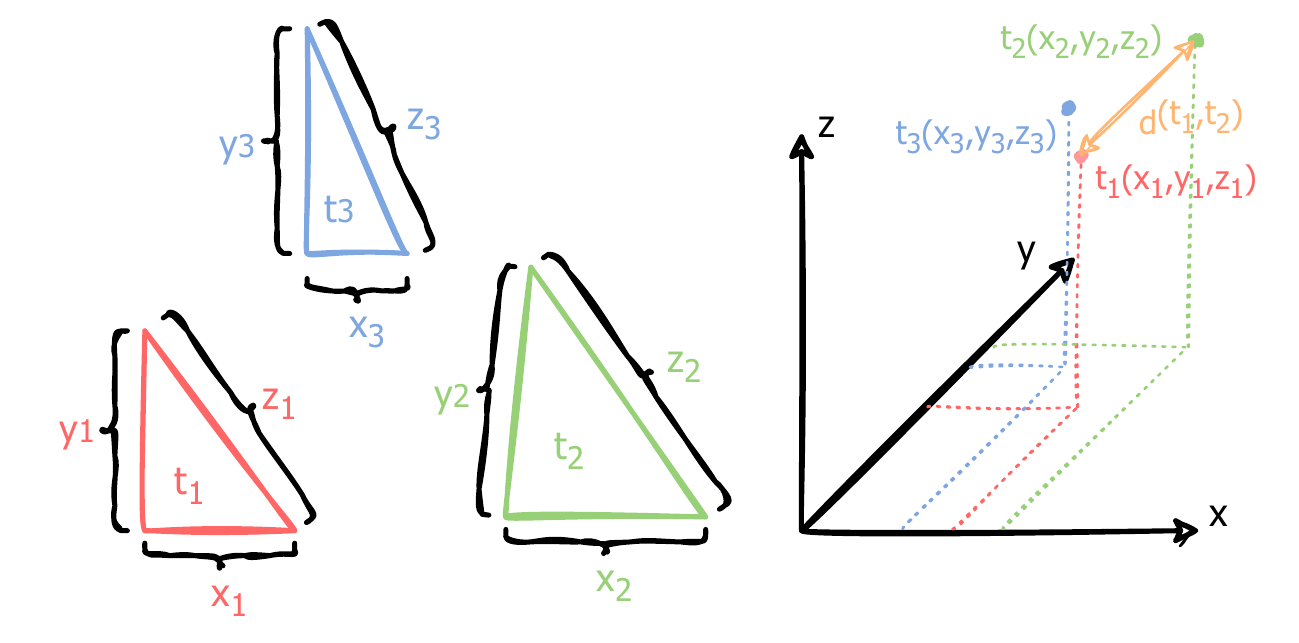}
    \caption{Triangle distance metric based on the sorted lengths of individual edges.}
    \label{fig:distance-metric}
\end{figure}
 
\subsection{K-means Clustering}
\label{sec:K-means}
We map the triangles into a 3D space according to the lengths of their edges (see \autoref{fig:distance-metric}). Each triangle $t_i$ is placed at point $p(x_i,y_i,z_i)$ such that $x_i,y_i$, and $z_i$ represent the length of shortest, middle, and longest edge of the triangle $t_i$ respectively. In this way, the position of each triangle in the 3D space shows the parameter of the triangle. Similarly, the Euclidean distance $d_{(i,j)}$ between any two triangles $t_i$ and $t_j$ represents their degree of similarity. The smaller the distance  more similar the triangles are, and vice versa. The two similar triangles will lie on the same point, yielding zero distance. Mathematically, the algorithm attempts to minimize the following energy function: 
\begin{equation}\centering
\label{main_eq}
 F_d(M_f)=\sum_{i=1}^{k} \sum\limits_{\substack{j=1\\t_j \in c_i }}^n d(t_j,t_i^*),
\end{equation}
where $F_d(M_f)$ is the energy function (accumulative distance), $k$ is number of clusters, $n$ is the total number of triangles/patches, $c_i$ is the $i^{th}$ cluster, $t_j$ is triangle in $c_i$, and $t_i^*$ is the centroid (mean) of cluster $c_i$, calculated as:  
\begin{equation}\centering
\label{centeralT_eq}
 t_i^*= \frac{1}{|c_i|}\sum_{t_j \in c_i} t_j.
\end{equation}
Similarly, $d(t_j,t_i^*)$ is the value indicating the degree of dissimilarity between $t_j$ and $t_i^*$, which is calculated as square Euclidean distance between their positions ($p(x_j,y_j,z_j)$ and $p^*(x_i^*,y_i^*,z_i^*)$, respectively) in our 3D space. Mathematically,  
\begin{equation}\centering
\label{sim_eq}
d(t_j,t_i^*)={|x_j -x_i^*|}^2+{|y_j -y_i^*|}^2+{|z_j -z_i^*|}^2.
\end{equation} 
In other words, $|x_j -x_i^*|$, $|y_j -y_i^*|$, and $|z_j -z_i^*|$ are the differences between the shortest, middle, and longest edges of the two triangles, respectively. An abstract view of our clustering scheme is presented in \autoref{alg:clustering}.


\begin{algorithm}
\caption{Clustering}
\label{alg:clustering}
\KwData{$M_f, k, T$}
\KwResult{$Labels, WithinClusterDistances$} 
  \begin{algorithmic}[1]
    \STATE $Matrix{[n\times n]} \gets Virtual3Dspace$
    \FOR{\textbf{each} $cell(i,j)$ of $Matrix{[n\times n]}$}
      \STATE $cell(i,j) \gets d(t_i,t_j)$ \quad // Dissimilarity between  $t_i$ and $t_j$ (calculated via \autoref{sim_eq}).
    \ENDFOR \textbf{\ each}
    \STATE $\left[Labels, WithinClusterDistances\right] \gets \kmeans(Matrix{[n\times n]},k)$
    \STATE Label each triangle according to its $Cluster$ number.
    \STATE \textbf{return} $\left[Labels, WithinClusterDistances\right]$
  \end{algorithmic}
\end{algorithm}

For clustering, we first create a $n \times n$ matrix where each cell $(i,j)$ contains the value of $d(t_i,t_j)$ calculated via \autoref{sim_eq}. Then, we apply K-means clustering to the matrix, which classifies the triangles into $k$ clusters, where the user specifies $k$. Each triangle is labeled with its value of $k$ representing its class. For each cluster, the central triangle is calculated via \autoref{centeralT_eq}. Then remeshing operators are applied to minimize the distance, defined with \autoref{sim_eq}, of any triangle of the cluster to its cluster's central triangle.

\paragraph{Convergence point and clustering error:} 
The convergence point of \autoref{alg:main} is reached when the maximum value of the within-cluster distances (\autoref{maxCError_eq}) reaches a given threshold. This maximum value indicates the highest difference among the triangles in the same cluster. We define our clustering error of any triangle as the Euclidean distance from its cluster's central triangle. The maximum error $(Error_{max})$ and mean error $(\overline{Error})$ can be calculated from \autoref{sim_eq} as follows. 
\begin{equation}\centering
\label{maxCError_eq}
Error_{max}= max(d(t_j,t_i^*)),  \forall 1\le i\le k, t_j \in c_i, 
\end{equation}
and, 
\begin{equation}\centering
\label{meanCError_eq}
\overline{Error}= mean(d(t_j,t_i^*)),  \forall 1\le i\le k, t_j \in c_i.
\end{equation}

\subsection{Remeshing Pipeline}
\label{sec:remesh}
Local operators (edge flip, edge collapse, and vertex translation) are used for surface remeshing (see \autoref{alg:remesh}). The objective of remeshing is to minimize the distances between the triangles in each cluster (\ie to make them similar). The objective is achieved by minimizing the energy function defined with \autoref{main_eq}. The algorithm works as follows: starting with a \ac{cvt}-initialized mesh, the algorithm first improves the regularity of vertices by making its valency (number of adjacent edges) optimal. Edge flip operators make the valencies of the vertices equal to or near 6 (a regular vertex). Regular vertices converge quickly during surface remeshing~\citep{Khan2022,Vidal2015}.

As stated in \autoref{sec:K-means}, the shape of each triangle is mapped into a single point in a virtual 3D space. The central triangle of each class of triangles (from surface mesh), which is calculated using \autoref{centeralT_eq}, is presented by a central point of the cluster (in the virtual 3D space). The distance from each triangle to the corresponding centroid is calculated using the degree of similarity between the two triangles (\autoref{sim_eq}). Making triangles similar in the surface mesh brings the corresponding points in 3D space closer to the central point. To minimize this distance, the algorithm applies direct remeshing operators, including edge flip, edge collapse, and vertex translation.

\paragraph{Edge flip and collapse:}
Remeshing algorithm (\autoref{alg:remesh}) is called from the main algorithm (\autoref{alg:main}). For the first iteration of the main algorithm, the edge flip and edge collapse operators are applied in \autoref{alg:remesh} if this improves the similarity values. Next is the vertex translation, which is applied in each iteration.


\begin{algorithm}
\caption{Remeshing}
\label{alg:remesh}
\KwData{$M_s, k, T, Itr$}
\KwResult{$M_f$ with k classes of identical triangles} 
  \begin{algorithmic}[1]
    \IF{Itr=1}
      \STATE $M_f \gets Valence Optimization(M_s)$
      \IF{CollapseImprovesClusteringDistance}
        \STATE $\mathrm{Collapse}()$
      \ENDIF
      \FOR{\textbf{each} edge $e_i \in M_f$}
        \IF{EdgeFlipImprovesClusteringDistance}
          \STATE $\mathrm{EdgeFlip}()$
        \ENDIF
      \ENDFOR \textbf{\ each}
    \ENDIF
    \FOR{\textbf{each} edge $e_i \in M_f$}
      \STATE Calculate required new length of $e_i$ \quad // \ie difference from cluster's center via \autoref{sim_eq}
      \STATE Calculate its pressure on both side vertices
    \ENDFOR \textbf{\ each}
    \FOR{\textbf{each} vertex $v_c \in M_f$}
      \STATE Calculate pressure from each adjacent edge
      \STATE Calculate new position $p^*$ \quad // \ie mean of the pressures from adjacent edges (see \autoref{vtranslat_eq})
      \IF{ShapeIsPreserved}
        \STATE Translate $v_c$ to new position $p^*$
      \ENDIF
    \ENDFOR \textbf{\ each}
    \STATE \textbf{return} $M_f$
  \end{algorithmic}
\end{algorithm}

\begin{figure}[tb]
    \centering
    \includegraphics[width=\linewidth]{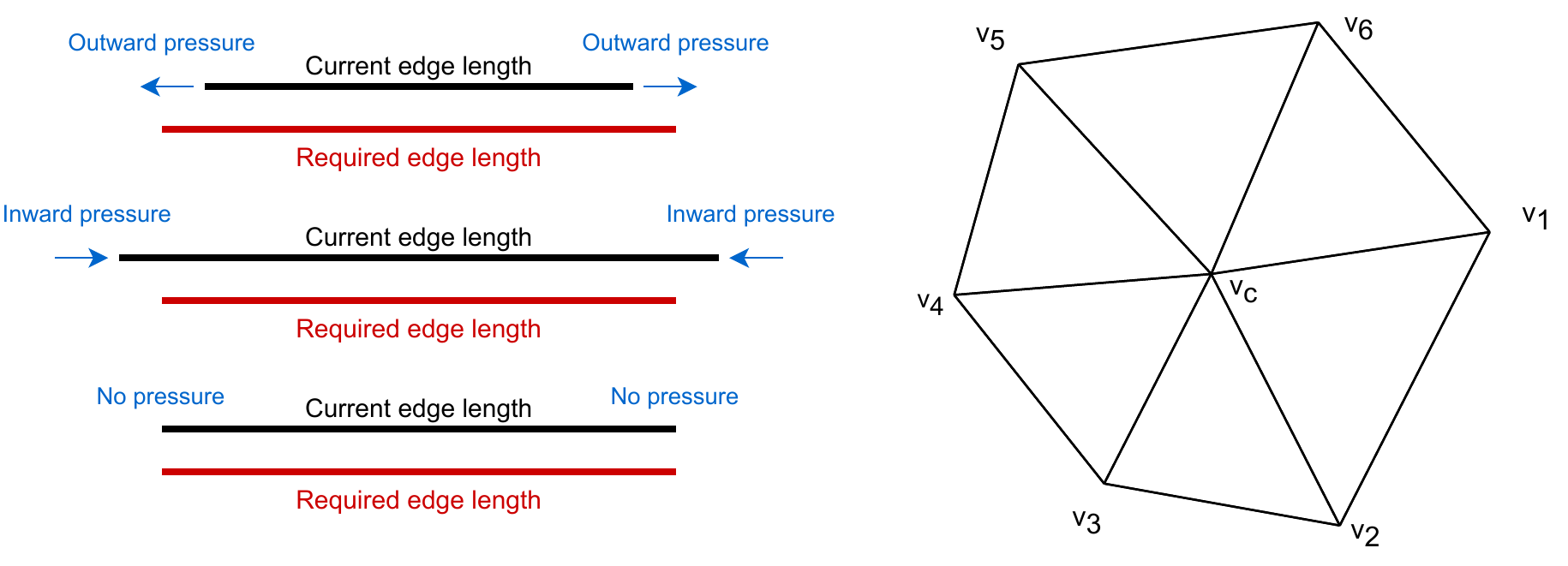} 
    \caption{Pressure calculation for vertex translation. Left: three cases of pressure direction. The current edge length is increased/decreased/kept unchanged depending on the required edge length. Right: 1-ring neighborhood of a vertex $v_c$ and calculation of accumulative pressure.}
    \label{fig:vtranslate}
\end{figure}
\paragraph{Vertex translation:}
Vertices are translated using \autoref{alg:remesh}. The algorithm requires information on whether to increase or decrease the edge length (see~\autoref{fig:vtranslate} (left)). \autoref{sim_eq} puts pressure on each edge, indicating the magnitude of change in edge length to make it similar to the corresponding edge of the cluster's central triangle. The pressure is divided on both side vertices of each edge. Since each vertex is connected to multiple edges, it is affected by pressure from different side edges (see~\autoref{fig:vtranslate} (right)). \autoref{fig:vtranslate} (right) shows 1-ring neighborhood of vertex $v_c$. Each vertex $v_i$ from the neighborhood of $v_c$ puts pressure on $v_c$, attempting to bring it to a new position  $p_i$. For a vertex with $n$ adjacent vertices, \autoref{vtranslat_eq} calculates the average of $n$ new positions to find the final position $(p^*)$ for $v_c$. The above steps are applied iteratively until the algorithm converges.
\begin{equation}\centering
 \label{vtranslat_eq}
 p^* = \frac{1}{n}\sum_{i=1}^{n} p_i.
\end{equation}
To ensure geometric preservation and avoid possible deformation, we put restrictions on vertex translation. The vertex translation is only allowed if new position $(p^*)$ is inside the predefined limit. In case it lies outside, the value of $p^*$ is recalculated as a middle point  between $p^*$ and the current position of the vertex. This limit is defined on each side of the surface. Typically, this limit is kept as $0.25^{th}$ of the mean edge length. However, for a quicker convergence, this value can be increased.

\subsection{Puzzle and Assembling}
\label{sec:puzzle-assembly}
\begin{figure}[!htb]
    \centering
    \includegraphics[width=0.8\linewidth]{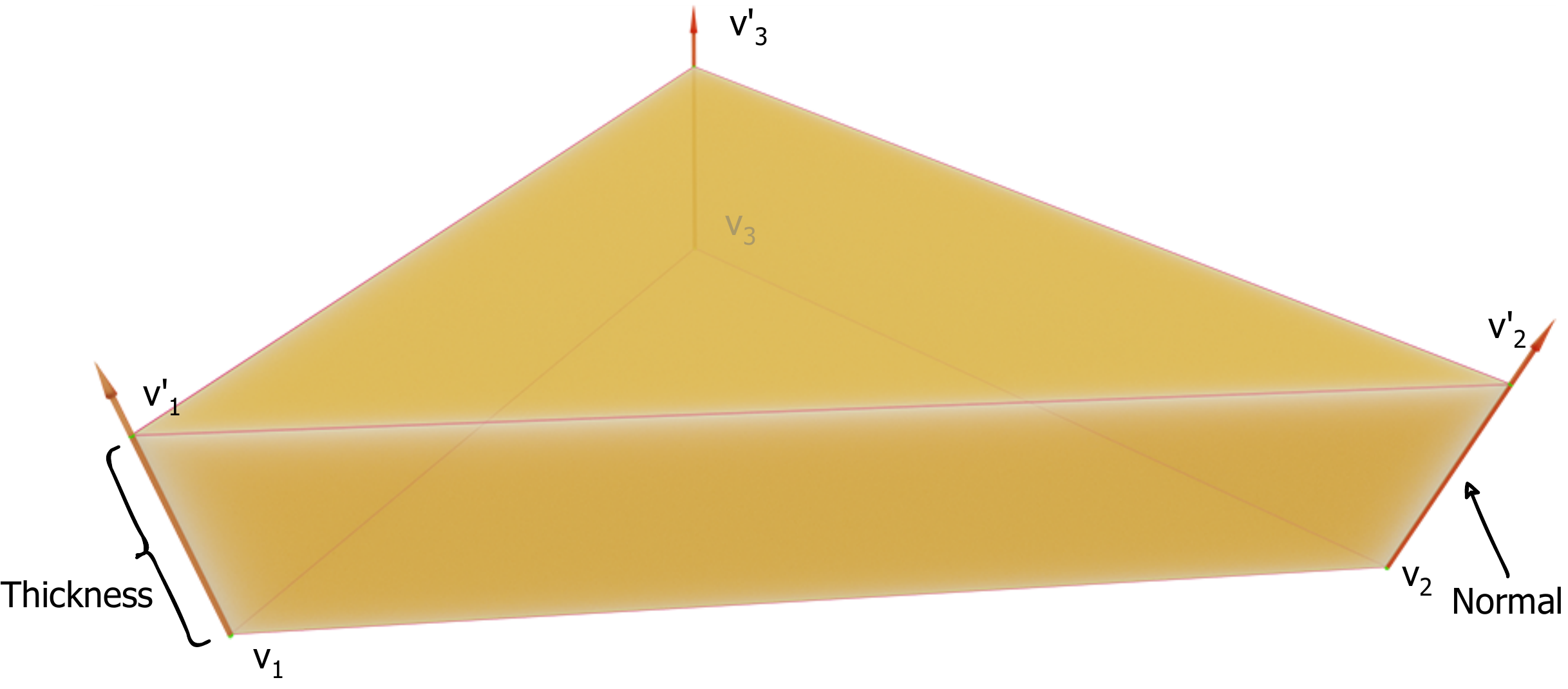} 
    \caption{The individual triangular patch is thickened along the vertex normals.}
    \label{fig:thickness}
\end{figure}

we aim to generate reconfigurable puzzle segments from the input surface. To ensure that segments are reconfigurable, we thicken the surface by adding a new surface layer and connecting it with the existing one resulting in a triangle patch with user-defined thickness as shown in \autoref{fig:thickness}. The thickened triangle patches are processed for female connections (holes), allowing interconnection with other segments via a hinge joint connector.

\paragraph{Thickening:}
For creating an external surface layer, we calculate the vertex normals (irrespective of its clustering). The vertex normal is calculated as the mean of the normals of the adjacent triangles. Next, we insert a new vertex ($v^{\prime}_1$, $v^{\prime}_2$, $v^{\prime}_3$) on the normal of each vertex at a distance of user-given thickness. We connect the corresponding vertices, which make quad patches at three sides of the thickened triangle (see \autoref{fig:thickness}). This thickened triangle patch is a closed mesh having two triangles \ie a bottom triangle ($v_1$, $v_2$, $v_3$) laying on the inner layer and an upper triangle ($v^{\prime}_1$, $v^{\prime}_2$, $v^{\prime}_3$) laying on the outer layer and three quads. Finally, for a unique thickened triangle for each cluster, we calculate the mean of possible variations in the vertices' normals.

 The new vertices ($v^{\prime}_1$, $v^{\prime}_2$, $v^{\prime}_3$) added for the thickness also define the curvature. If we use  the vertex normal, the outer layer ($v^{\prime}_1$, $v^{\prime}_2$, $v^{\prime}_3$) of each part (thickened triangle) will be bigger, equal, or smaller than inner one ($v_1$, $v_2$, $v_3$) depending upon its position on the surface, which can be concave, planar or convex. Face normal, on the other hand, yields similar sizes for both layers without regard to curvature. 

\begin{figure}[!htb]
\centering
\begin{subfigure}{0.326\textwidth}
\includegraphics[width=0.9931\linewidth]{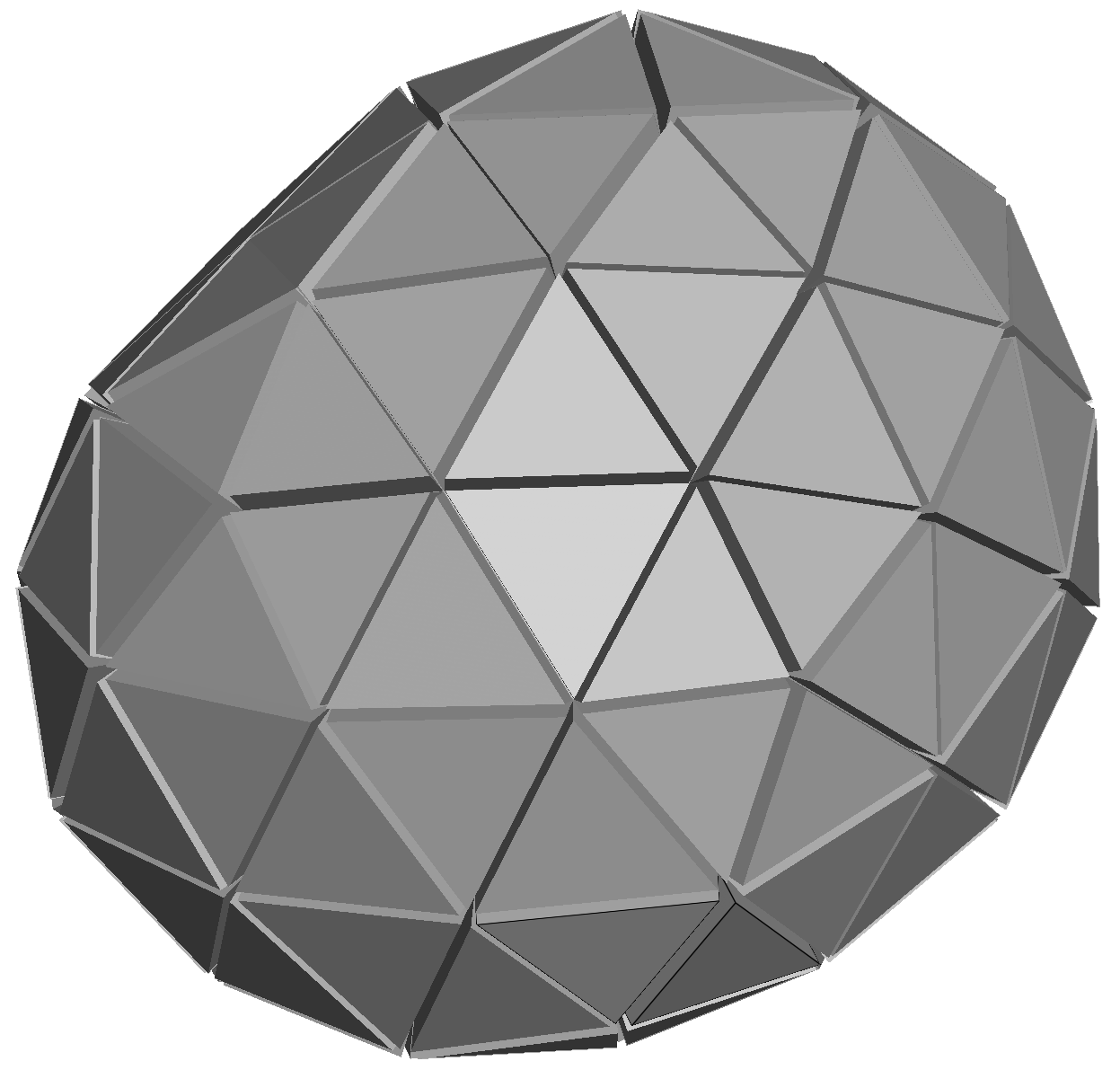}
    \caption{Perpendicular extrusion.}
    \label{sbfig-thk-A}
\end{subfigure}
\begin{subfigure}{0.326\textwidth}
\includegraphics[width=0.9931\linewidth]{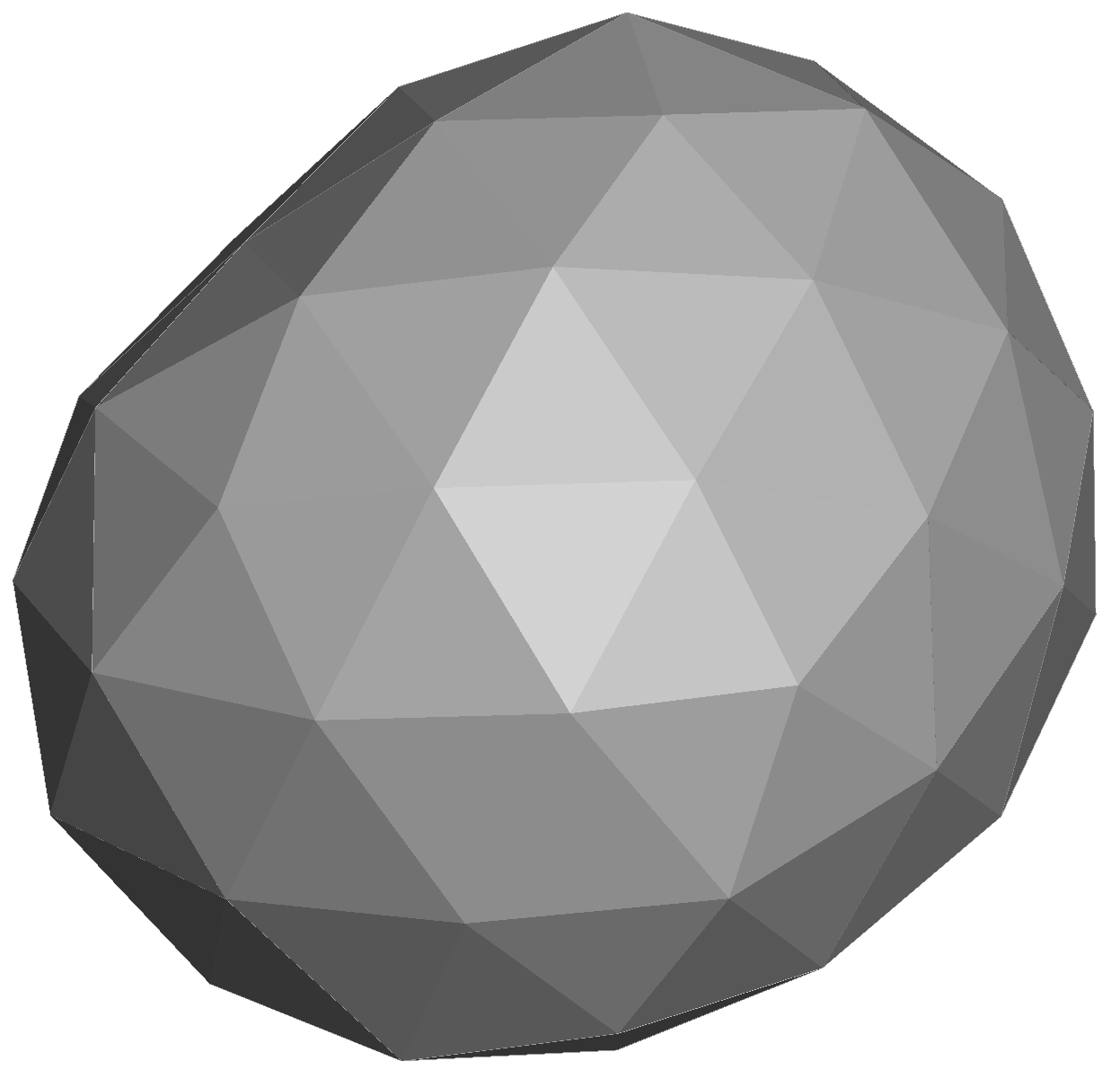}
    \caption{Curvature-aware }
    \label{sbfig-thk-B}
\end{subfigure}
\begin{subfigure}{0.326\textwidth}
\includegraphics[width=0.9931\linewidth]{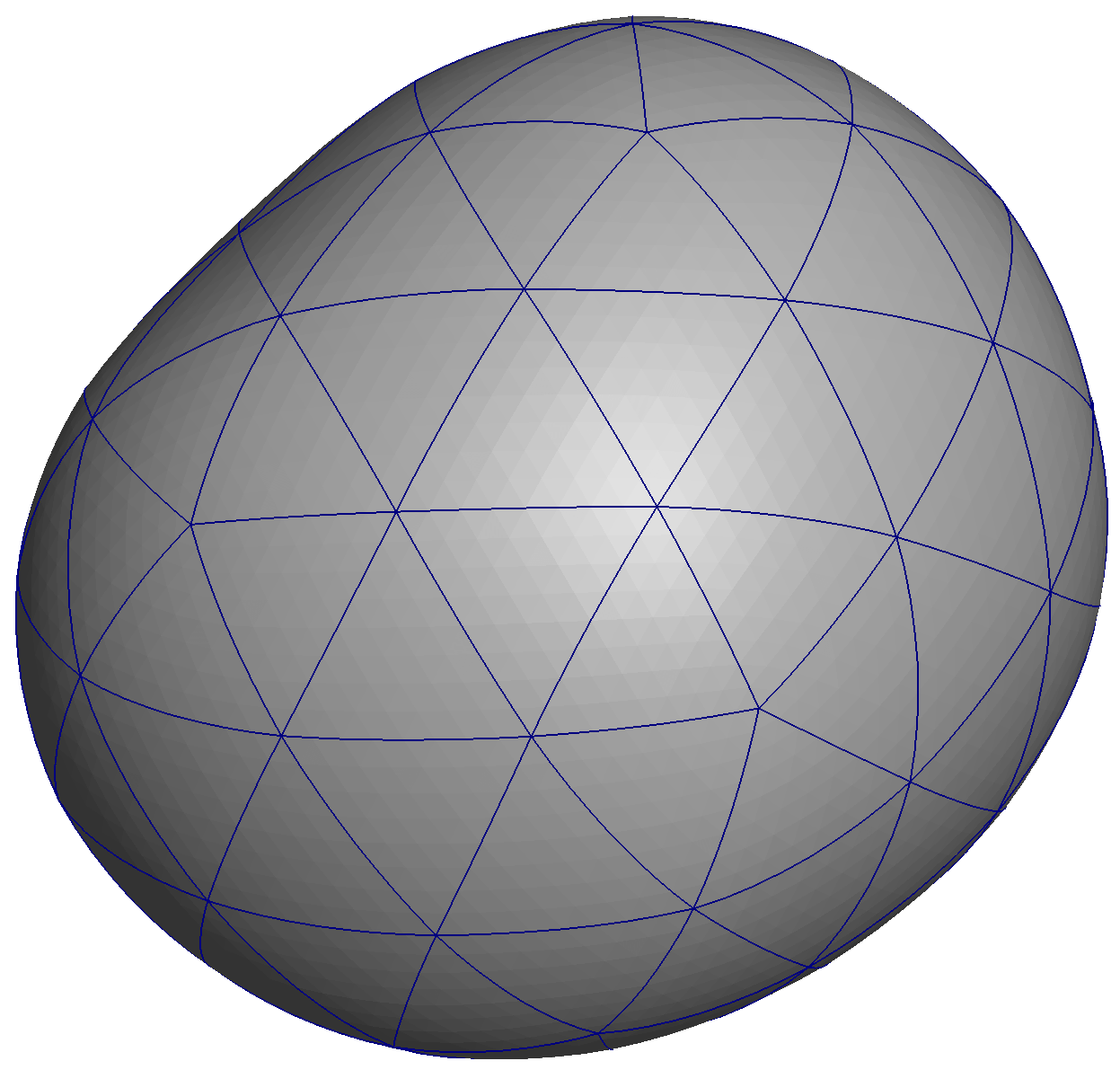}
    \caption{After smoothing.}
    \label{sbfig-thk-C}
\end{subfigure}
   \caption{Thickness layer: \subref{sbfig-thk-A}: The outer layer is created via face normal. \subref{sbfig-thk-B}: The outer layer is created via the vertex normal (Curvature-aware extrusion). \subref{sbfig-thk-C} Smoother with subdivision.}
   \label{fig:3thickness}
\end{figure}

\autoref{fig:3thickness} shows the thickened triangle patches using different approaches (see \autoref{sbfig-thk-A} and \ref{sbfig-thk-B}) on a full surface. In the case of face normals (perpendicular extrusion), for each triangle patch, the inner and outer layers are the same; therefore, there is some free space between them in the composed model. We use the vertex normal instead, which enables us to fill those spaces. Since the outer layer is dependent on curvature, the thickened triangles from the same class might have different sizes of the outer layer. To handle this issue, we find mean curvature for all parts in each class to finalize the positions of the new vertices ($v^{\prime}_1$, $v^{\prime}_2$, $v^{\prime}_3$). The three quads on the sides of the triangle are further processed for holes.

Subdivision smoothing gives even better results (see \autoref{sbfig-thk-C}). The inner and outer triangle surface patches are linked with quad strips linking the corresponding vertices. The number of vertices along each edge is defined by the level of the applied subdivision smoothing. These thickened triangle patches are further processed for connection holes before it is ready to be printed.

\begin{figure}[!htb]
\begin{center}
\includegraphics[width=0.8\linewidth]{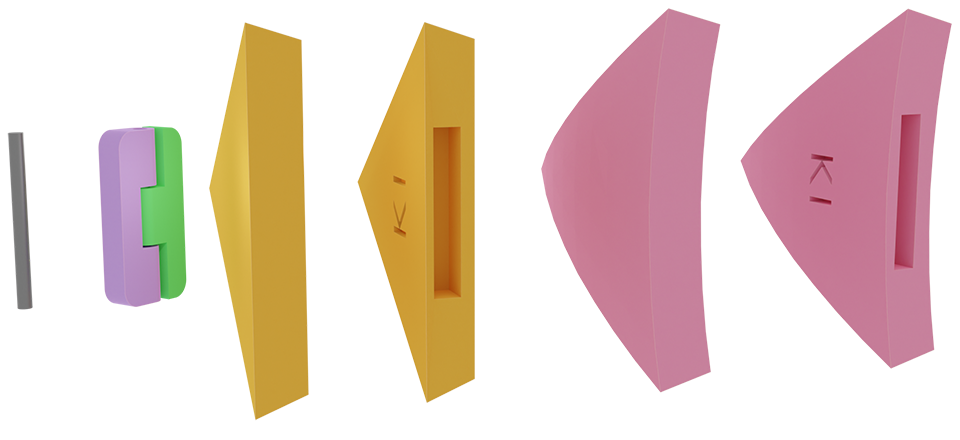}
\vskip -0.3cm
   \caption{Puzzle parts: A thickened triangle with and without holes and a hinge joint connector.}
   \label{fig:puzzle-parts}
\end{center}
    \vspace{-5mm}
\end{figure}

\paragraph{Connecting structures:}
Since the thickened triangle patches' sides differ significantly between the planar and curved triangle patches, we designed separate procedures for generating connecting parts for them. The connecting structures consist of holes in the thickened triangular patches and hinge joint connectors (see \autoref{fig:puzzle-parts}).

\begin{figure}[!htb]
\begin{center}
\includegraphics[width=1.0\linewidth]{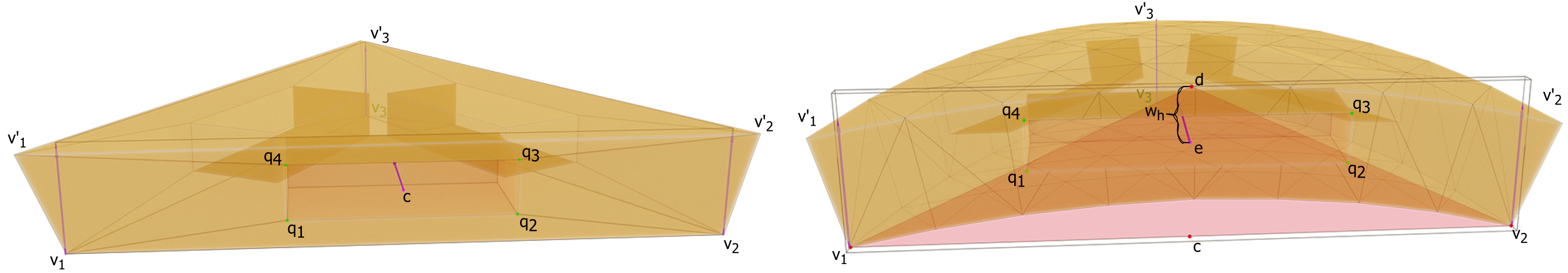}
\vskip -0.13cm
   \caption{Outline of hole generation process for connecting neighboring planar (left) 
 and curved (right) triangle patches.}
   \label{fig:Holes}
\end{center}
    \vspace{-5mm}
\end{figure}

\noindent \textbf{Planar triangular patch:}
To preserve the structural strength of individual pieces and provide some clearance tolerance for the connecting structures, we create holes parallel to the inner and outer surfaces. To achieve this, we calculate the center point ($c$) of the quad extruded from vertices (\eg $v_1$ and $v_2$) perpendicularly to the inner surface touching the outer surface (see top image in \autoref{fig:Holes}). We define the position of vertices $q_1, \ldots, q_4)$, which determine the hole boundary, according to the size of the hinge joint connector defined by the user. Next, we extrude the quad defined with vertices $q_1, \ldots, q_4$ inwards according to the size of the hinge joint connector. We remesh the outer side surface by connecting original vertices $v_1, v_2, v_3, v^{\prime}_1 v^{\prime}_2, v^{\prime}_3$ with newly created vertices $q_1, \ldots, q_4$. The top image in \autoref{fig:Holes} outlines the process. The same process is repeated for all three sides of the triangular patch.

\noindent \textbf{Curved triangular patch:}
Creating holes in the curved triangular patches is a bit more challenging. We first align the triangular patch so that the vertices $v_1, v_2, v_3$ lie on the XY-plane. Next, we calculate a projection point $d$ of the midpoint between vertices (\eg $v_1$ and $v_2$) $c$ to the outer edge of the same side (see the bottom image in \autoref{fig:Holes}). We define the hole center point $e$ at the half-thickness distance $w_h$ from point $d$ to $c$. We define the positions of vertices $q_1, \ldots, q_4$, which determine the hole boundary, with respect to the center point $e$ and the size of the hinge joint connector,  defined by the user, on a plane defined with vertices $v_1, v_2, d$. The quad defined with vertices $q_1, \ldots, q_4$ is then extruded inwards according to the size of the hinge joint connector. The side surface of the thickened curved triangular patch is remeshed by incorporating the vertices $q_1, \ldots, q_4$. The same process is repeated for all three sides of the triangular patch.

\noindent \textbf{Hinge joint connectors:}
The holes are identical for all triangles. Similarly, considering the box-like shape of the created holes defined by eight vertices, we generate appropriate hinge joint connectors for all edges to connect triangles with each other. The assembled hinge joint connector fits inside two holes (with some spacing to compensate for the printer inaccuracies). Its size is determined by the user, who must also consider the possible overlaps of the holes within the individual triangular patch if they are too big. The two parts of the hinge joint connector are printed with a resin material (similar to triangles). However, inside the connector is a metal rod (see \autoref{fig:puzzle-parts}).

\subsection{Flexibility and extendability}
We mainly generate puzzle parts from planar triangular patches. However, we also implemented the support for curved triangles. Similarly, our technique also provides the option of merging triangles into other polygonal structures, which reduces the puzzle's complexity if intended by the puzzle designer.

\paragraph{Sub-division and Smoothing } 
\label{sec:subdivision}
The mesh complexity has been reduced during the initialization step of our algorithm. However, the simplified mesh has a rigid surface with discrete planar triangles. To improve surface smoothing, we follow the approach by~\cite{Loop1987} to smooth the mesh by subdivision of triangles. The curved triangles are easy to assemble and provide a smoother surface. However, it is very hard to improve their isometric decomposition. Therefore, we only create curved triangles and apply classification with a higher number of $k$.

\paragraph{Curvature-aware patch classification: }
With subdivision, each planar triangle patch is divided into 64 triangles (in 3 iterations $4^3$), making a curved patch. We store the boundaries of each patch and calculate its curvature. Next, we apply patch-wise classification, which applies not only the parameter but also the curvature. For curvature-aware patch classification, we add a $4^{th}$ dimension $w$ to \autoref{sim_eq} as follows:
\begin{equation}\centering
\label{sim_eq_curved}
d'(t'_j,t_i^{\prime *})={|w_j -w_i^*|}^2+{|x_j -x_i^*|}^2+{|y_j -y_i^*|}^2+{|z_j -z_i^*|}^2,
\end{equation}
where $t'_j$ and $t_i^{\prime *}$ are the $j^{th}$ curved triangle in $i^{th}$ class, and center of the cluster, respectively. To calculate $w$, we find the center of the planar triangle (formed by three corner vertices of the curved triangle) and the center of the central sub-triangle of the subdivided curved triangular patch and calculate the square Euclidean distance between them. The curvature-aware patch classification gives smoother results. However, if two patches have an abrupt change from concave to convex (or vice versa), the uniform width calculation for patch thickness becomes challenging.

\paragraph{Merge Triangles} 
For further flexibility, the user can also choose to merge triangular patches into convex polygonal patches (full or parts of the pentagons and Hexagons). Depending upon user choice, we provide different merging options, including two, five, six, or seven triangles. The merging strategy can minimize the puzzle's complexity.

\section{Experimental Results}
This section evaluates \drkid on different organic shapes such as SARS-CoV-2 virions from the Electron Microscopy Data Bank\footnote{\url{https://www.ebi.ac.uk/emdb/EMD-33297}} under id EMD-33297~\citep{Ngan2022}, mitochondria from the UroCell Dataset~\citep{ZerovnikMekuc2020,ZerovnikMekuc2022}, and cell nuclei from the WTC-11 hiPSC Single-Cell Image Dataset v1~\citep{Viana2023}. We show some assembled models of these shapes and compare them with their 3D models in \autoref{fig:print}. The proposed algorithm is implemented in C++ and tested on  Intel(R) Xeon(R) Gold 6230R CPU  $2\times 2.10$ GHz with 156 GB RAM and $64$ bit Windows 10 operating system.

\subsection{Validation}
\begin{figure*}[!htbp]
\begin{center} 
\includegraphics[width=0.997\linewidth]{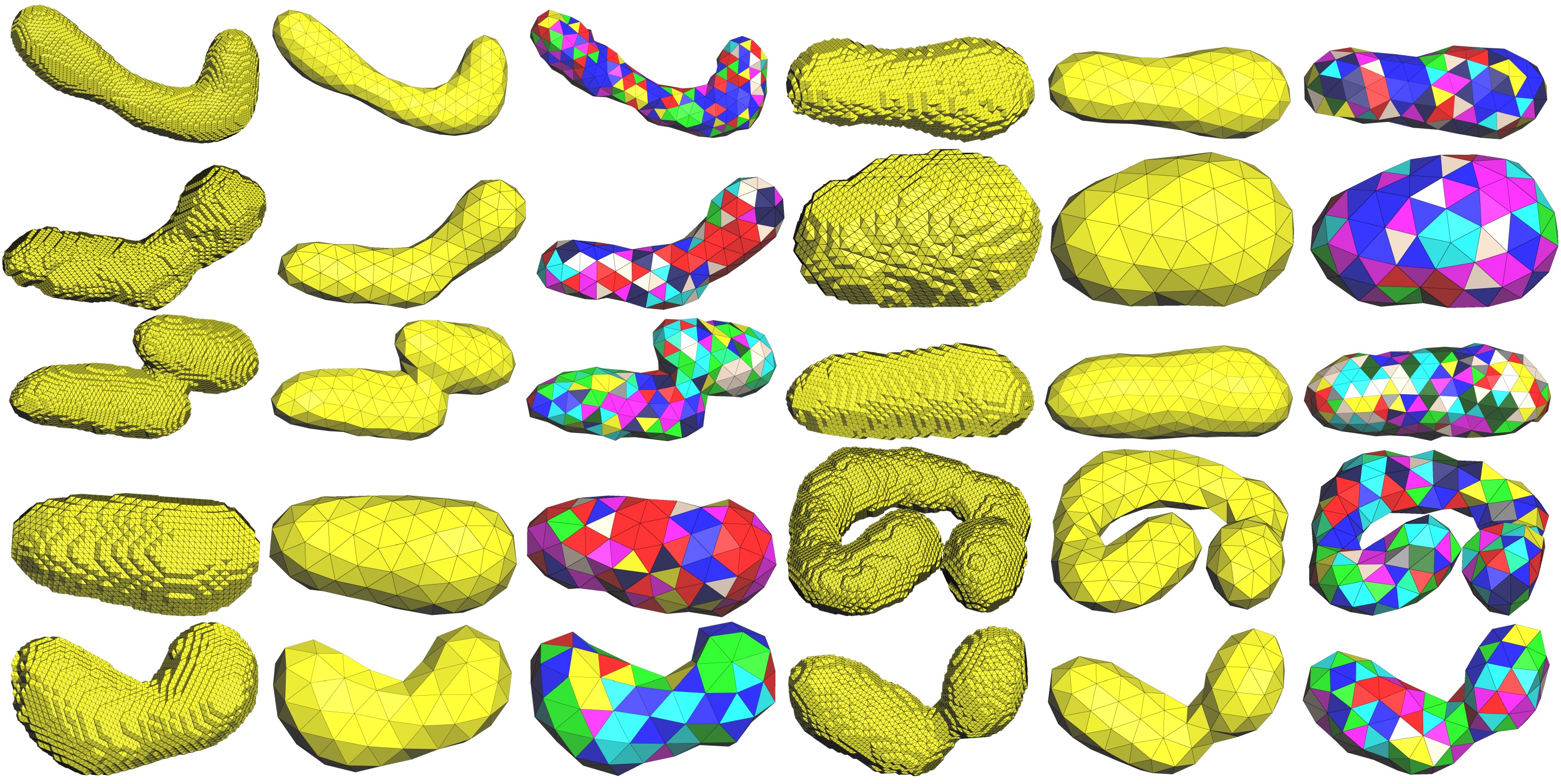} 
   \caption{Clustering results of 10 different models. From top left to bottom right, we name the models $M1$ to $M10$. For each model, the three sub-figures (from left to right) are mesh generated from the 3D shape ($M_i$), the simplified mesh ($M_s$), and our results ($M_f$). The corresponding numerical results are given in \autoref{tab1e:stat}.} 
   \label{fig:mainRzlt}
\end{center}
\end{figure*}

\begin{figure}[!htbp]
    \centering
    \includegraphics[width=1.0\linewidth]{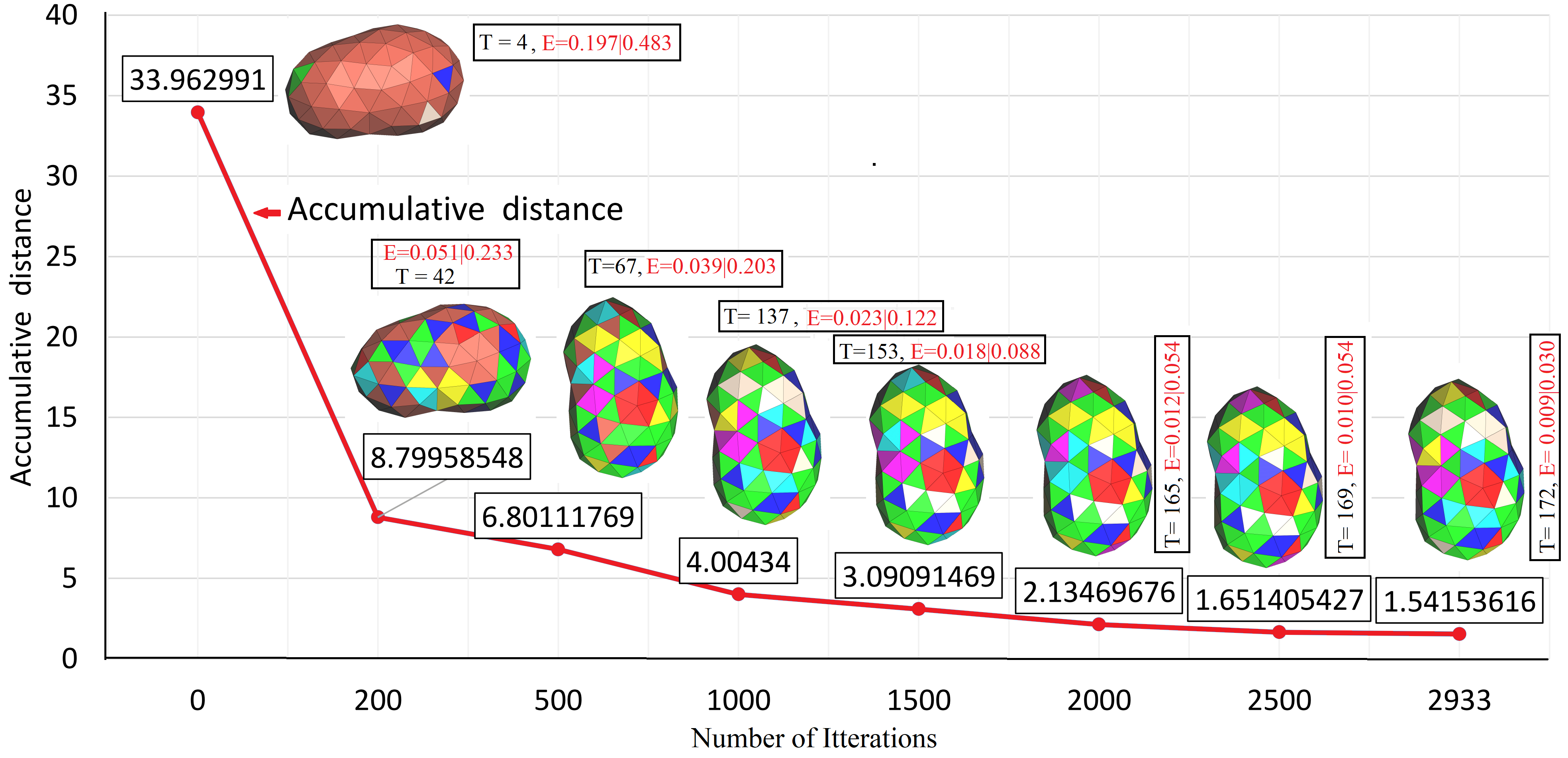} 
    \caption{Minimization of the accumulative distance (\ie energy function~\autoref{main_eq}) with number of iterations. Here $T$ presents the number of triangles with error smaller than the threshold (0.03). $E$ represents the clustering error (mean|max). The brown-colored triangles have errors above the threshold.}
    \label{fig:m4cplot}
\vskip -0.53cm
\end{figure}

\begin{figure}[tb]
\begin{center}  
\includegraphics[width=1.0\linewidth]{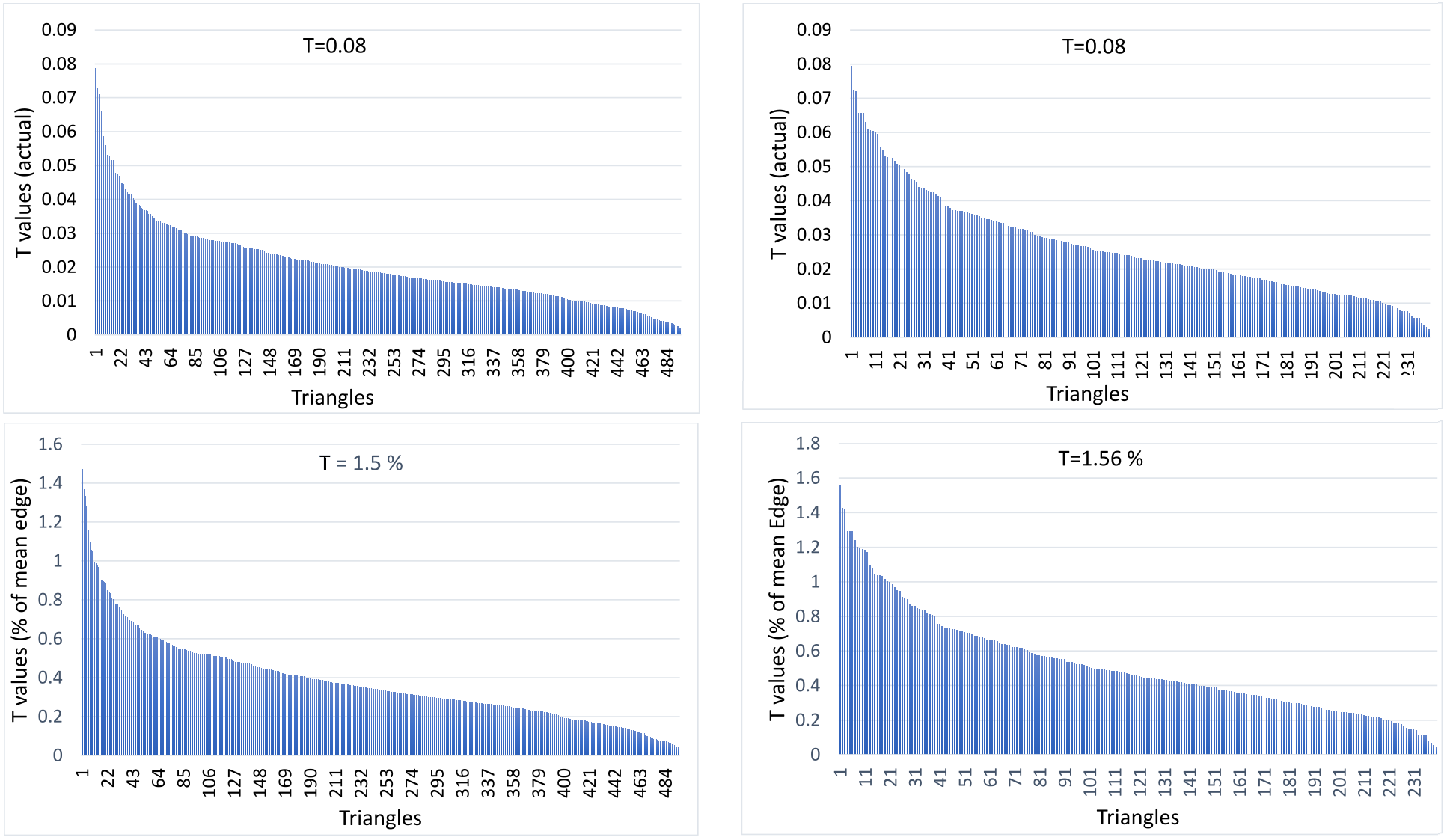} 
   \caption{Histogram of clustering error for two models. Top actual values and bottom percent values. Left: M1 with an average error of $0.38\%$ for 496 triangles. Right: M2 with an average error of $0.53\%$ for 240 triangles. The values are calculated as a percentage of the mean edge length. Here x-axis represents the triangle number, whereas the y-axis shows the clustering error (distance from the cluster center in $\%$ of mean edge length).} 
   \label{fig:Histograms}
\end{center}
\end{figure}

In this section, we present empirical results to validate the applicability and effectiveness of our algorithm. We calculated the actual cluster error and a relative error, \ie percentage of the mean edge denoted as $\%(e)$. The relative error gives an easy judgment of the accuracy as it considers the actual length. Typically, the remeshing algorithms and isometric decomposition need only negligible (acceptable) geometric loss. We calculate the geometric loss in the percentage of bounding box value of the Hausdorff distance~\citep{Barton2010}. As stated in \autoref{sec:algoOverview}, our algorithm generates a raw mesh ($M_i$), which is preprocessed and simplified to $M_s$. The simple mesh $M_s$ is then iteratively remeshed consecutive with \kmeans clustering to reach a given threshold. Additionally, there may be some geometric loss during mesh simplification $d_H(M_i,M_s)$ and in clustering/remeshing $d_H(M_s,M_f)$. \autoref{fig:mainRzlt} shows the clustering results for different models. \autoref{tab1e:stat} shows quantitative results, including the geometric loss, efficiency, classification error, and the number of iterations. Observing the numerical results, we can conclude that the geometric loss during the clustering is smaller than the geometric loss in simplification. However, there is a trade-off between the clustering threshold $T$ and geometric loss ($d_H$). The threshold $T$ given to the algorithm shows the limit of acceptable clustering error. Since we are dealing with variable-sized models, the $T$ shown in the paper is the percentage of the mean edge length. The algorithm iterates until the errors are under the threshold limit. Therefore, the average clustering error is smaller than $T$ (see \autoref{fig:m4cplot}, and \autoref{fig:Histograms}).

The algorithm is iterative and proceeds toward its convergence by minimizing the clustering error and the value of the accumulative energy function. \autoref{fig:m4cplot} plots the energy minimization (\autoref{main_eq}) and shows the convergence of the algorithm with the number of iterations. The clustering errors also decrease as the algorithm iterates. 


\subsection{Printing Results}
The prototypes were printed using two 3D printers: Stratasys J750, a high-resolution full-color 3D printer\footnote{\url{https://support.stratasys.com/en/Printers/PolyJet-Legacy/J735-J750}} and FormLabs Form3 SLA 3D printer\footnote{\url{https://formlabs.com/3d-printers/form-3/}}. All the hinge joint connectors and most of the models were printed using Stratasys 3D printer. Only triangles of the big SARS-CoV-2 virion membrane model were printed using the FormLabs printer. The printing resolution and precision of the FormLabs printer are not good enough for the final model to be assemblable. These 3D fabrication results are prototypes validating the plausibility of our method. The envisioned production will use the molding process, where our low demand on the number of required molds will secure the production scalability.

\begin{figure}[!htbp]
\vskip -0.3cm
\begin{center}   
\includegraphics[width=\linewidth]{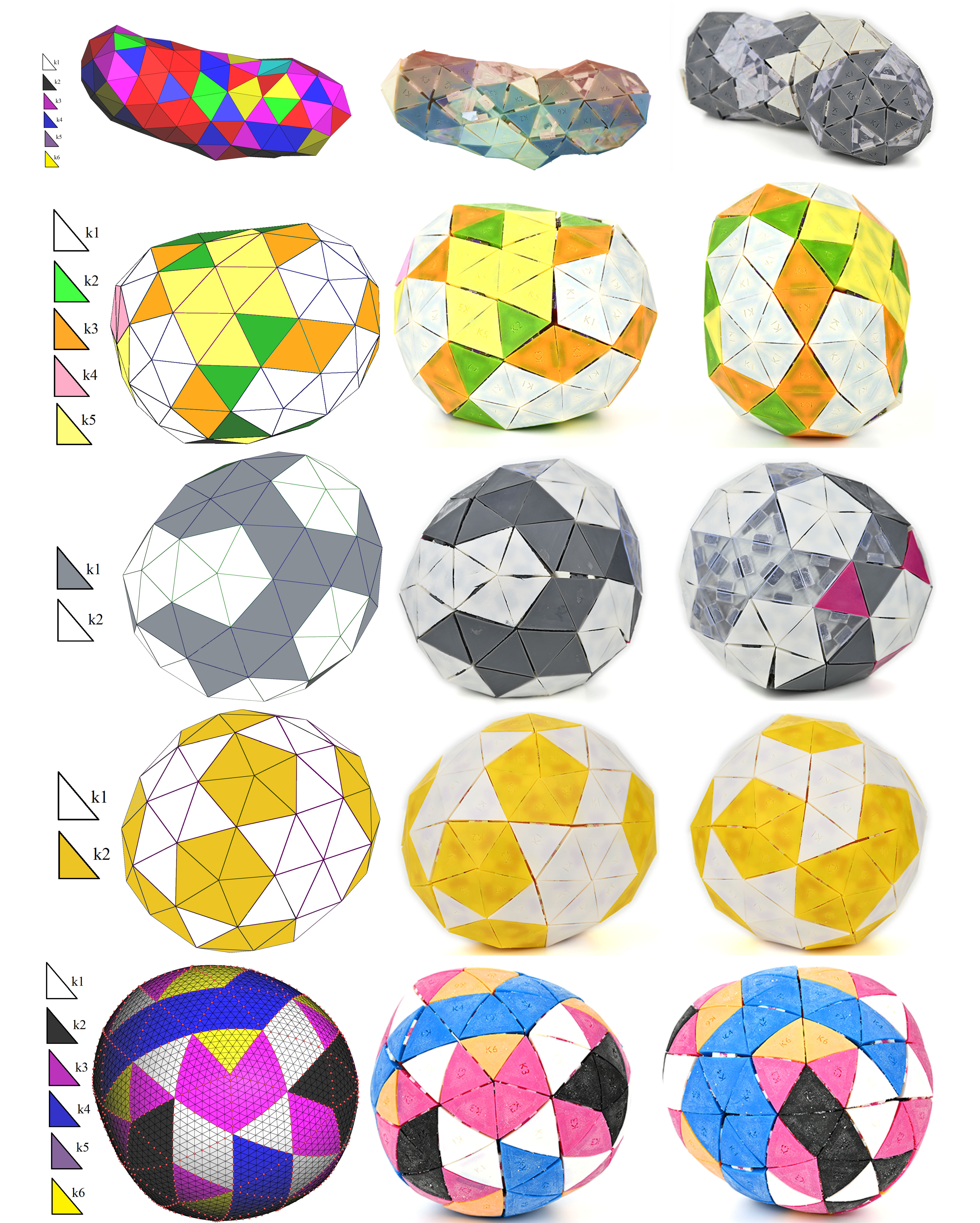}
\vskip -0.13cm
   \caption{Printed results. Top to bottom: Mitochondria outer membrane, cell nuclei membrane, and three models of SARS-CoV-2 virion membranes. The bottom row shows a model with curved triangle patches. The labels on the left column indicate the values of $k$ and segment colors. The number of colors of the printed models in the first and third row is smaller than the number of classes due to the limited colors available for 3D printing.} 
   \label{fig:print}
\end{center}
\vskip -0.5cm
\end{figure}

The 3D-printed results of several models are displayed in \autoref{fig:teaser} and \ref{fig:print}. \autoref{fig:print} also shows how the 3D models relate to their 3D-printed versions. We show that the printed results are similar to the original models. However, the tight connectors (hinge joints) are difficult to use for creating reconfigurable joints during the assembly phase. Therefore, we used loose hinge joints, which create gaps near the vertices with valence $\ge 7$. The size of the gaps also depends on the printer's accuracy.

\begin{figure}[!htbp]
\begin{center}    
\includegraphics[width=0.34\linewidth]{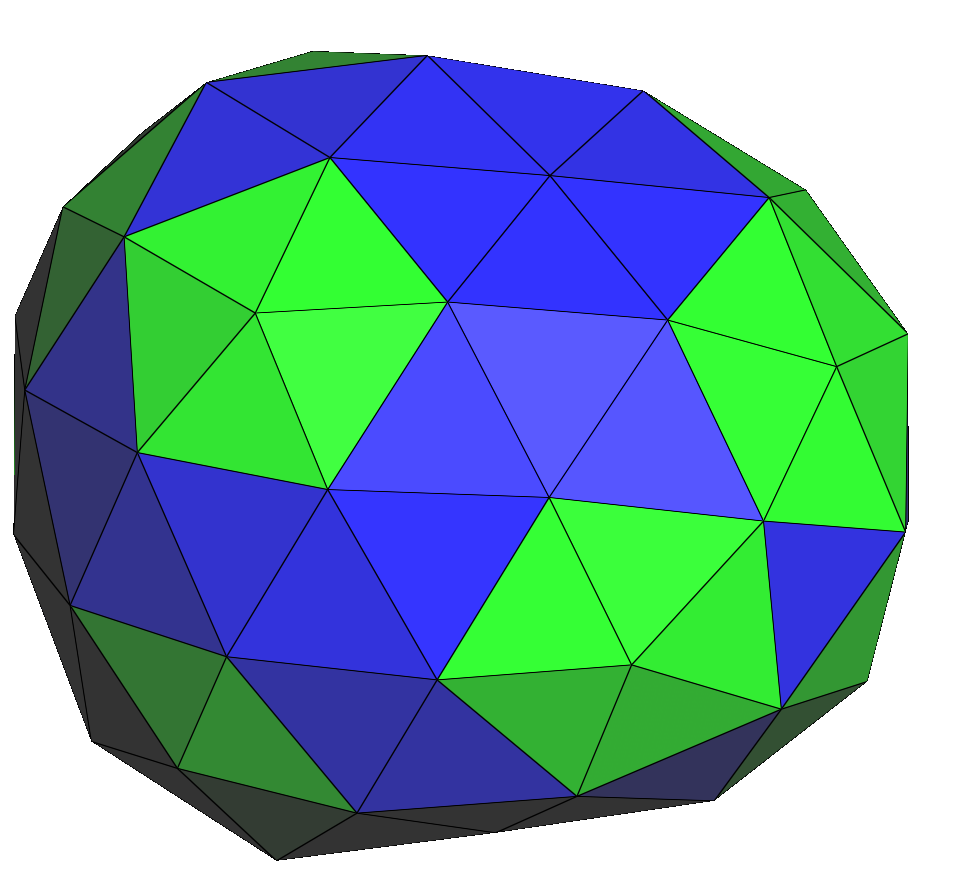}\hfill
\includegraphics[width=0.323\linewidth]{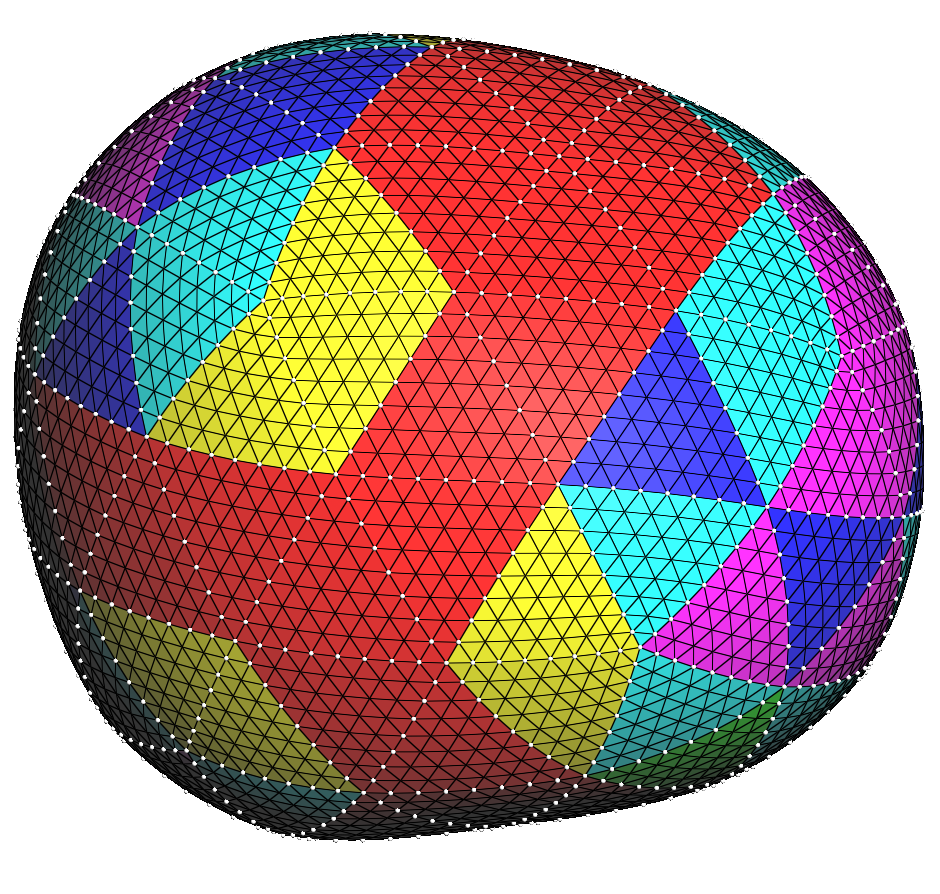}\hfill
\includegraphics[width=0.323\linewidth]{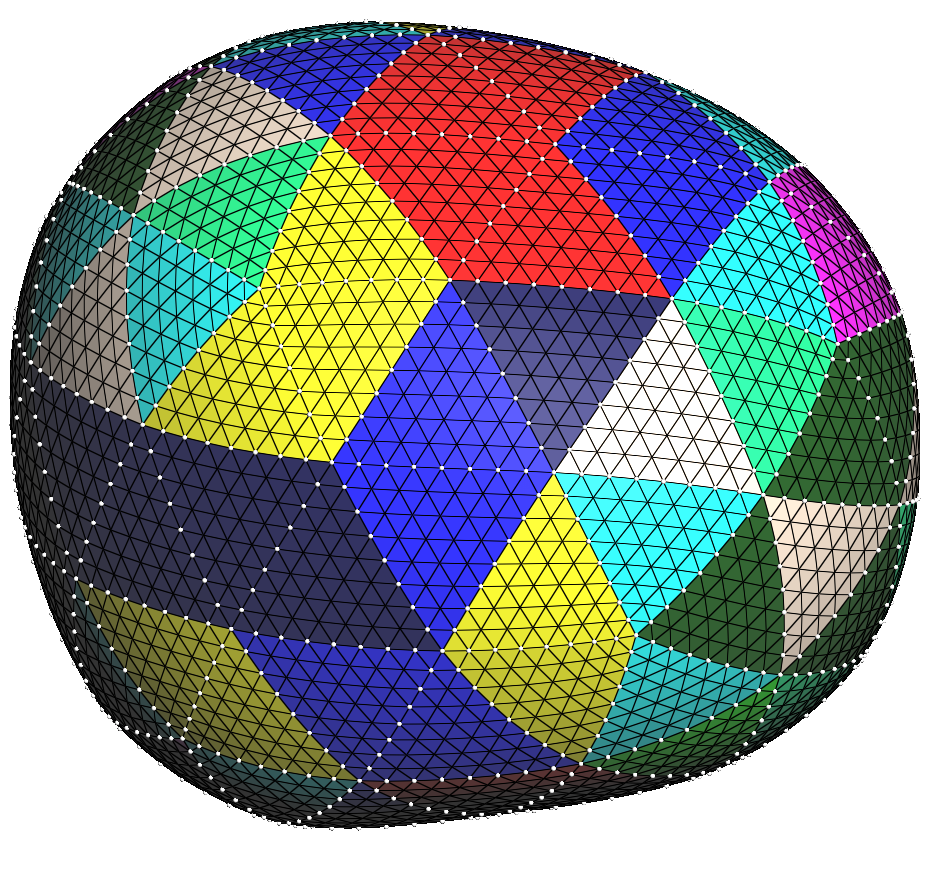}
   \caption{Patch-wise classification: Left: Clustering of planar triangles with $k=2$, $T=0.01\%$, Middle: Patch-wise classification with $k=6$, $T=7.50\%$, Right: $k=10$, $T=3.75\%$ } 
   \label{fig:smoothermodels}
\end{center}
\end{figure}

\paragraph{Classification of the curved patches:} In addition to planar triangles, \drkid can be used for curved triangular patches. \autoref{fig:smoothermodels} shows the results of a model composed of curved triangular patches. For curved triangular patches, we only provide the classification of the patches without any further remeshing. To reach a suitable classification, users must find an optimal combination between $T$ and $k$. For example, the model in \autoref{fig:smoothermodels} can classify all the patches (116) with $T=7.5\%$ if $k=6$. However, this patch-wise classification can be achieved with a lower threshold $T=3.75\%$ if we set $k=10$. The last row in \autoref{fig:print} shows a 3D-printed model with curved triangular patches.

\subsection{Comparison with related methods}
\begin{figure}[!htbp]
\begin{center}
\includegraphics[width=0.99440\linewidth]{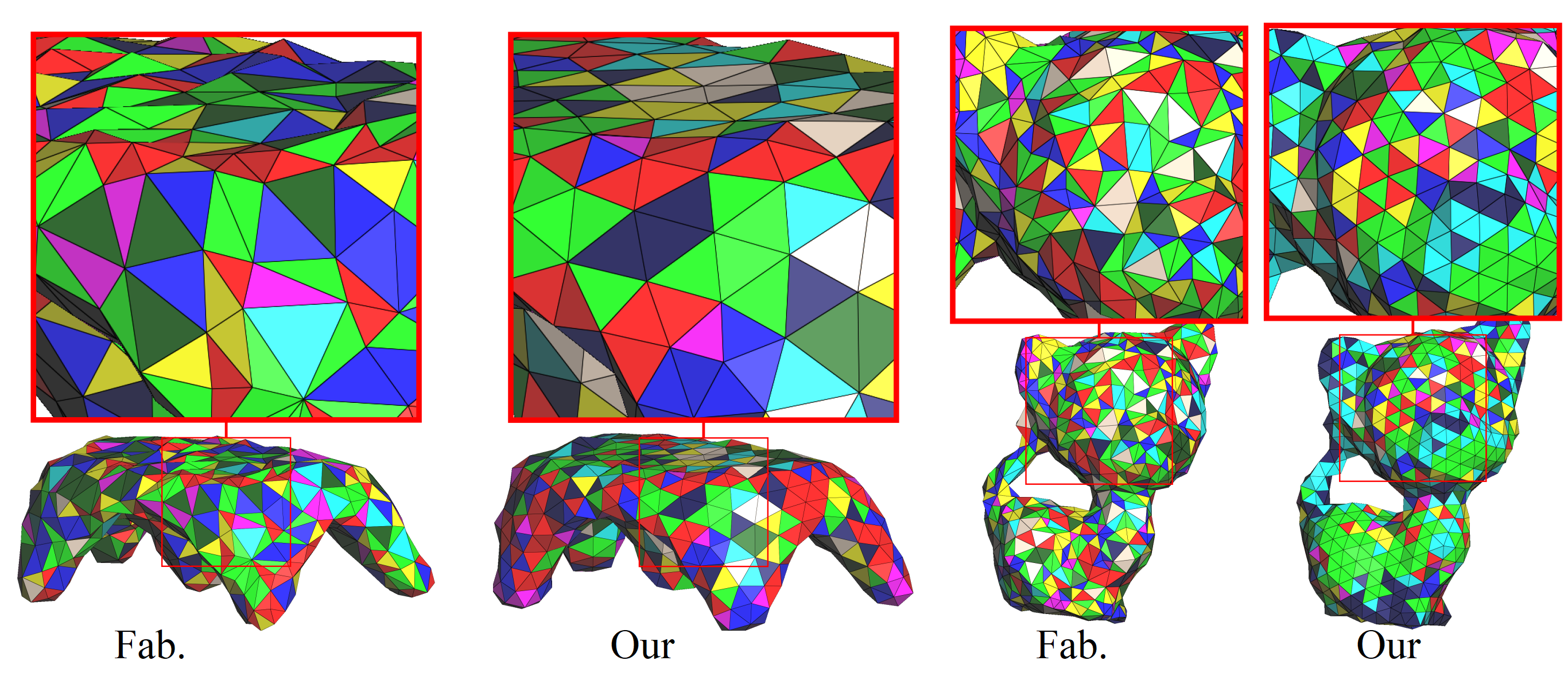} 
   \caption{Comparison with a previous method~\citep{Liu2021} for \textit{k=9}. Left (in each pair): Fabrication method~\citep{Liu2021} with $d_H(\%bb)=1.62$ and $1.33$, and with time taken 85.6 minutes, and 91.2 minutes. Right (in each pair): Our with $d_H(\%bb)=1.53$ and $1.31$, and with time taken 38 seconds and 27 seconds.  }
   \label{fig:compirision}
\end{center}
\vskip -0.5cm
\end{figure}

In the domain of biological structures, there is no standard algorithm that can be used for comparison with our approach. However, there are few articles in other domains. We choose a most recent article~\citep{Liu2021} for a short comparison using two models.  \autoref{fig:compirision} shows the comparative results. The results show that our method gives smoother results with a lower $d_H$. Moreover, our method is faster.

\begin{table*}[tb]
\caption{Quantitative results. Here $M_i$ is the input mesh, $M_s$ is the simplified mesh, and $M_f$ is the resulting mesh of our method. $T$ is user given threshold, and $d(t,t^*)$ is the average Euclidean distance from the cluster centers. Here, $\overline{Length(e)}$ is the mean length of edges, $\%(e)$ is the percent value of the mean of $\overline{Length(e)}$, $d_H$  is calculated in $\%$ bounding box. $\overline{Error}$ is the mean clustering error calculated via~\autoref{meanCError_eq}.}
\label{tab1e:stat}
\tabcolsep 3pt 
\begin{center}
\resizebox{0.999\textwidth}{!}{
\begin{tabular}{|l|c|c|c|c|c|c|c|c|c|c|c|c}
\hline
 Figure \#/Model&\# Faces&\# Vertices&k&T &$\overline{Length(e)}$&$\overline{Error}$&$d_H(M_i,M_s)$&$d_H(M_s,M_f)$ &$d_H(M_i,M_f)$ &Time& Iterations \\ 
 &$M_i/M_s/M_f$&$M_i/M_s/M_f$& &actual/$\%(e)$& &actual/\%(e)&mean/max   & mean/max  &mean/max&Sec.& \#\\ 
\hline
\autoref{fig:mainRzlt}/ M1&21280/500/496&10642/252/250&7&0.08/1.50&5.33&0.020/0.382&1.33/2.55&0.62/1.77 &3.38/1.25 &248&7893 \\

 \autoref{fig:mainRzlt}/ M2&10460/242/240&5232/123/122&9&0.08/1.57&5.08&0.026/0.519         &2.60/4.33  &0.49/1.87      &2.46/4.44 &15&993 \\
\autoref{fig:mainRzlt}/  M3&14048/264/262&7026/134/133&8&0.06/1.03&5.79&0.018/0.317         &1.61/3.62  &0.65/2.16      &1.50/3.72 &41&2385 \\
\autoref{fig:mainRzlt}/ M4&5968/172/172&2986/88/88&7&0.08/1.53&5.20&0.025/0.475           &0.50/1.47   &0.72/3.27      &0.83/2.71 &15&1546 \\

\autoref{fig:m4cplot}/ M4b&5968/172/172&2986/88/88&7    &0.03/0.57&    5.21&0.009/0.172&    0.50/1.47     &0.83/3.65        &0.91/3.06   &42&2933 \\
\autoref{fig:mainRzlt}/  M5&16072/374/372&8038/189/188&11&0.09/1.56&5.74&0.028/0.490      &0.22/0.95    &0.50/2.50      &0.51/2.05 &59&2349 \\
 \autoref{fig:mainRzlt}/ M6&9612/416/416&4808/210/210&11&0.09/2.17&4.14&0.032/0.775      &0.23/0.71     &0.83/3.15      &0.98/3.45&48&1962 \\
 \autoref{fig:mainRzlt}/ M7&7280/196/196&3642/100/100&8&0.08/1.44&5.57&0.023/0.416       &0.39/1.00     &0.90/3.37      &0.88/3.48 &33&2894 \\
\autoref{fig:mainRzlt}/ M8&21288/400/396&10646/202/200&11&0.09/1.39&6.49&0.023/0.355            &0.29/0.90     &0.67/2.83      &0.65/2.71 &101&5134 \\
\autoref{fig:mainRzlt}/ M9&10380/156/156&5192/80/80&6    &0.07/0.98&    7.12&0.023/0.320&     0.39/1.07      &0.90/3.20&0.84/3.58    &33&3170 \\
\autoref{fig:mainRzlt}/ M10&11496/196/196&5750/100/100&8    &0.08/1.20&    6.68&0.035/0.520&    0.36/1.72    &0.67/2.38         &0.72/2.57&17&1933 \\
\autoref{fig:print}/top&4807/172/170&2406/88/87&6    &17.3/1.89&    914&6.968/0.762&   0.34/1.23    &0.967/3.20        &0.85/2.34     &22&987 \\
\autoref{fig:print}/N-cell&83040/136/136&41522/70/70&6    &0.50/1.70&    29.32&0.180/0.613&    1.33/0.49    & 1.24/2.94         &2.11/4.17     &51&4732 \\

\autoref{fig:smoothermodels}/left&320/116/116&162/60/60&2    &0.02/0.01&    160&0.007/0.004&   0.50/1.04     &1.10/3.32        &1.93/4.70     &144&8349 \\
\hline
\end{tabular}
}
\end{center}
\end{table*}

\section{Discussion}
\label{sec:discussion}
\drkid is the novel system for the physicalization of organic shapes. It can be an effective tool for learning objects' physical properties and structure by providing a physical shape. The physicalization starts from a 3D shape and ends with a physical model assembled from reconfigurable parts. We provide a simple physical structure of small biological structures so users can assemble and see the basic structure. An abstract view of the structure is provided to the audience as a physical model, whereas the detailed functionalities of these models are not explored. We aim to attract the audience by providing a simple reconfigurable model. The models can also play a vital role in scientific museums or awareness conferences for public outreach.

The key goals of \drkid are: (1) to reduce cost by creating isometric segments, (2) to provide a puzzle-like reconfigurable physical model, and (3) to preserve the shape and curvature of the input shape.

For the isometric decomposition, we triangulate the surface mesh and design an energy function to increase the degree of similarity of triangles belonging to the same class. We proposed a simple yet novel method of computing similarities between triangles. Previous methods~\citep{Liu2021,Singh2010,Fu2010} used the vertex distances of two aligned triangles, so they need optimization to find the best alignment transformation. On the other hand, our method does not require this optimization of finding the best alignment transformation and is, therefore, computationally efficient.

\autoref{fig:m4cplot} shows the minimization of this energy by increasing the similarities among the triangles. For this energy minimization, we used surface remeshing, which can preserve the input shape. \autoref{fig:compirision} shows that our method provides smoother and more accurate results than the existing method. For puzzle-like assembly, we provide an automatic thickness and hole creation strategy. The thickness is created in the direction of vertices' normals and thereby preserves the curvature during assembly. If there is a small error in the curvature estimation, it can be compensated for by the freedom of hinge joint rotation.

The isometric decomposition provides a cost-effective solution for physicalization using injection molds or 3D printing, the former especially if a large quantity is demanded. Similarly, for a small number of prints, since we need to process each patch for holes and hinges, the isometric patches are easy to handle, where we process only one patch from each class of isometric patches.

The holes and hinge joints provide a connection between the triangles. The hinge joints can provide a certain degree of freedom to recover the possible geometric loss/error in curvature during the puzzle parts generation. In other words, the angle between the two faces (triangles) can be increased/decreased due to the use of hinge joints. \autoref{fig:print} shows our printed results which show a similar visual result to the input model.

The algorithm converges when the degree of similarities among the triangles reaches its threshold. In other words, if the clustering error goes below an error threshold. Typically, the parameters, including threshold (T), the k value, and the $d_H$ have a trade-off. Selecting a smaller value of $T$ will increase the $d_H$ or require a higher value of $k$. If we keep $d_H$ adjustable by the algorithm and set higher values of $T$ and $k$, the algorithm can reach convergence with a relatively smaller value of $d_H$. However, no mechanism exists to automate the optimal balance among these three parameters. After convergence, the maximum error cannot exceed a given threshold in each class of similar triangles. For example, \autoref{fig:Histograms} shows that only a few triangles are close to the threshold.

To support extendability and give smoother results, \drkid uses triangle subdivision followed by the classification of the curved triangle patches (\autoref{fig:smoothermodels}). For curved patches, we used \autoref{sim_eq_curved}. However, this assumption of curvature estimation may fail in special cases such as complex models.

The scope of our study is limited to a small set of microbes, \ie tiny models having potato-like curved surfaces (\eg mitochondria, lysosome, endosomes, cellular nuclei, virus particles, \etc) without sharp features. Complex structures such as models with sharp edges, models with abrupt changes from concave to convex (for example,  brain surface), and generic CAD models are out of the scope of our algorithm. \drkid can provide a baseline for several interesting future research directions. For example, modeling realistic biological structures with more detailed information.
\section{Conclusion}
\label{sec:conclusion}
We have presented \drkid, a new physicalization approach for potato-shaped biological structures. \drkid generates the surface mesh of the model, which is then decomposed into $k$ types of identical triangle segments. We used a novel mapping function that maps the degree of similarity among the triangles into a virtual 3D space as a distance metric. Next, \kmeans clustering is applied to classify the triangles into $k$ classes. We used surface remeshing to make all the triangles in the same class similar. The segments are automatically generated as puzzle parts and thickened with hinge joints and female connections to support a reconfigurable assembly. We demonstrate how our approach can be used for 3D printing the prototype models. \drkid is considering a cost-reduction approach for the physicalization of small microbes. We hope it to be a practical and useful tool for the community.

Our approach has the following limitations:
\begin{itemize}[nolistsep]
    \item  We have evaluated our system empirically with different models, however, we have no theoretical proof of the success in isometric decomposition with an arbitrary model and a given number of $k$, however, the hinge joint connectors compensate for a slight imperfection in the fit.
    \item  Although for our current domain, since the target organic shapes are in several variations, we believe that the geometric loss is acceptable. However, for generic use, the geometric loss should be further considered.  
    \item  We only present a short comparison with another fabrication method. However, we did not find any other relevant physicalization algorithms for the biological domain for comparison. 
\end{itemize}

For the future, we are working on designing a specialized mesh generation algorithm to improve accuracy. We also plan to minimize geometric loss during remeshing and extend the algorithm's scope into other domains. Additionally, we aim to fully automate the process for hinge joint connectors without requiring user-defined dimensions. Furthermore, remeshing the curved patches (\autoref{fig:smoothermodels}) to make them isometric is also a future challenge. Finally, we aim to address other parts of the structures apart from membranes.

\section*{Acknowledgment} 
\vskip -0.14cm
This research was supported by King Abdullah University of Science and Technology (KAUST) (BAS/1/1680-01-01), the KAUST Visualization Core Lab, and the VCC Center Competitive Funding (CCF). 
The authors would like to thank Prof.\ Helmut Pottmann for the valuable discussion on the project and Prof.\ Xiao-Ming Fu for sharing the results of the fabrication method~\citep{Liu2021}. We would also like to thank the KAUST Prototyping Core Lab staff for their assistance in printing the models.
\FloatBarrier

\bibliographystyle{unsrtnat}

\bibliography{references}
\end{document}